\documentclass{llncs}
\usepackage{makeidx}  
\usepackage{graphics}

\usepackage{amsmath}
\usepackage{tensor}
\usepackage{multirow}
\usepackage{url}

\usepackage{rotating}
\usepackage{pdflscape}

\usepackage[sort]{cite}

\usepackage{pgf,tikz}
\usetikzlibrary{trees}

\begin{document}
%
%
\addtocmark{Math Search for the Masses: Multimodal Search Interfaces and Appearance-Based Retrieval} 

%
\title{Math Search for the Masses:\\Multimodal Search Interfaces and Appearance-Based Retrieval\thanks{The final publication is available at {\tt http://link.springer.com.}}}

\titlerunning{Math Search for the Masses}  
%
\author{Richard Zanibbi \and Awelemdy Orakwue}
\authorrunning{Zanibbi \& Orakwue} 
%
%
\institute{Document and Pattern Recognition Lab\\Department of Computer Science, Rochester Institute of Technology\\Rochester, NY 14623, USA}
\maketitle              

\begin{abstract}
We summarize math search engines and search interfaces
produced by the Document and Pattern Recognition Lab in recent years, and in particular the~$m_{in}$ math
search interface and the \emph{Tangent} search engine. Source code for both systems are publicly available.
``The Masses" refers to our emphasis on creating systems for mathematical
non-experts, who may be looking to define unfamiliar notation, or browse
documents based on the visual appearance of formulae rather than
their mathematical semantics. 
\end{abstract}
\section{Introduction: Why Math Search Pertains to the Masses}

Mathematical notation is a \emph{natural} language used to define the models, metrics and analytical tools of modern societies. It is natural in the sense that the notation is re-purposed and adapted for different mathematical concepts, problems, and communities, leading to various dialects. The influence of mathematical notation, while quiet, is pervasive. Whether it is choosing foods to purchase based on their cost and quantified nutritional information, or using demographic and usage statistics to determine which forms of entertainment to attract and promote, where to build manufacturing sites, or how to represent a problem and its potential solutions in science and technology, math notation is an essential tool that shapes both our personal lives and environment.  Given this, math literacy is critical for participating fully in the modern world, and considerable attention continues to be focused on strengthening mathematics education. 

However, for many persons of all ages, mathematical notation is a source of significant frustration or anxiety at times due to real or perceived difficulties with interpreting unfamiliar notation. This is particularly true when the text accompanying mathematical notation is found to be confusing. To search the internet for alternative sources about the notation, users must formulate their query in text, even if they are unclear what about what the represented concept is.
Mathematical experts might search the internet using \LaTeX~for an expression, or using the (often, already known) name for the concept \cite{zhaokan2008}. Even experts relate to the odd experience of revisiting concepts expressed in a notation distinct from that used when they originally learned a concept, and the difficulties this introduces in interpreting the notation.

While mathematical concepts can often be notated various ways, some psychological studies suggests that the appearance of math notation affects our reasoning about it \cite{Landy2007cj}, and that individuals will often space subexpressions systematically when entering math, even if mathematically unnecessary \cite{Landy2007lo}. The studies suggest that our perception of math notation may be grounded in visual structure, i.e. how it looks.

An important related concern is hit content summarization, i.e. how search hits are presented to the user \cite{youssef2007}. In a recent study it was confirmed that as one expects, the formatting of math expressions significantly affects relevance assessment of search hits \cite{Reichenbach2014}. The normal hit format provided by Google (e.g. as raw \LaTeX~or Presentation MathML) was compared with the same hits with formulae rendered,  and on average participants had a 17\% higher relevance assessment accuracy in the rendered condition. 

We propose that if it is natural to use words from unclear texts in queries, it is also natural to use mathematical notation from unclear texts \emph{directly} in queries. A recent study illustrates the benefit of this approach \cite{Wangari2014}. When undergraduate students were asked to learn about the binomial coefficient, and shown the expression $4 \choose 2$, many did not know what the notation represented. When given an interface in which they could draw, recognize the spatial layout of symbols and then search for the expression, most participants found this to be both intuitive and helpful.

In the remainder of our paper, we summarize research carried out to realize the entry and retrieval of math based on its appearance, as this is what one works from when notation is unfamiliar or difficult to interpret, or when trying to locate similarly structured expressions (e.g. when browsing formulae in a collection).

\section{Math Encodings: Symbol Layout Trees and Operator Trees}

In practice, math encodings are most commonly used to represent the appearance   
and mathematical content of formulae. Appearance-based encodings such as
\LaTeX~\cite{lamport1986document} and Presentation MathML\footnote{\url{http://www.w3.org/Math/}} are used to display mathematical expressions. A number of web browsers support Presentation MathML directly (e.g. Firefox), and online tools such as MathJax\footnote{\url{https://www.mathjax.org/}} may also be used to render \LaTeX~and MathML contained in HTML pages.  Mathematical content
encodings such as Content MathML can be used for evaluation and symbolic manipulation by
Computer Algebra Systems (e.g. Mathematica, Maple).

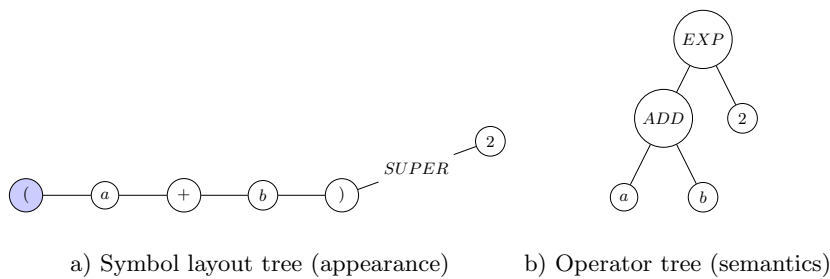
\begin{figure}[h]
\begin{tabular}{c c}
\scalebox{0.7}{\mbox{
\begin{tikzpicture}[grow=right]
\tikzstyle{every node}=[circle,draw]
\node[circle,fill=blue!20] {$($}
		child {
			node {$a$}
			child {
				node {$+$}
				child {
					node {$b$}
					child {
						node {$)$}
						child [grow=20] {
							node [draw=none,fill=none] {$SUPER$}
							child {
								node{$2$}
							}		
						}
					}
				}
			}
		};
\end{tikzpicture} } } &

\scalebox{0.7}{
\mbox{
\begin{tikzpicture}[grow=down]
\tikzstyle{every node}=[circle,draw]
\node {$EXP$}
	child {
		node {$ADD$}
		child {
			node {$a$}
		}
		child {
			node {$b$}
		}
	}
	child {
		node {$2$}
	};
\end{tikzpicture}}}
\\
~\\
a) Symbol layout tree (appearance)
&
b) Operator tree (semantics)
\end{tabular}
\caption{\it Symbol Layout Tree and Operator Tree for $(a+b)^2$ }\label{fig:trees}
\end{figure}

As illustrated in Figure \ref{fig:trees}, the appearance and content of mathematical expressions are hierarchical. As a result, both appearance and content-based encodings define trees, which are annotated in various ways to support applications. Encodings for appearance represent a spatial arrangement of symbols on baselines (writing lines), which we term a \emph{Symbol Layout Tree}. In Figure \ref{fig:trees}a the symbol layout tree is rooted at the leftmost parenthesis (`('), and there are two writing lines present: the main baseline, and a superscripted baseline.

The \emph{Operator Tree} in Figure \ref{fig:trees}b represents a hierarchical application of operations to operands, from the leaves to the root of the tree. Relative to the layout tree in Figure \ref{fig:trees}a, in the operator tree operator symbols are replaced by their associated operation  (e.g. `+' becomes $ADD$), implicit operations are made explicit (e.g. superscript becomes $EXP$ (exponent)), and parentheses used for grouping in the expression appearance are removed. Groupings are redundant in an operator tree, where ordering constraints are explicit.

Due to symbols and spatial relationships in formulae being frequently redefined by authors, it is impossible to define a mapping from formula appearance to formula semantics. To create this mapping the domain of discourse (e.g. algebra vs. calculus vs. logic, etc.) along with the specific environment defining constants, variables, and operations in an expression are required. 
The mapping from operator trees to layout trees is also one-to-many, as a single operator tree may be written various ways. For example, `$x^2$' and `$(x)^2$' can represent the same operation,  and operations may be associated with different symbols (e.g. `$\div$' vs. `$\slash$' for division) and symbol arrangements (e.g. $1/2$ vs. $\frac{1}{2}$). 

The flexibility of mathematical notation benefits both authors and their technical communities. However, this flexibility and dependency on context for interpretation poses substantial challenges for automated recognition and retrieval of mathematical notation \cite{zanibbi2012recognition,handbook2014}.

\section{$m_{in}$: a Multimodal Math Search Interface}
\label{sec:min}

\begin{figure}
\centering
\scalebox{0.15}{\includegraphics{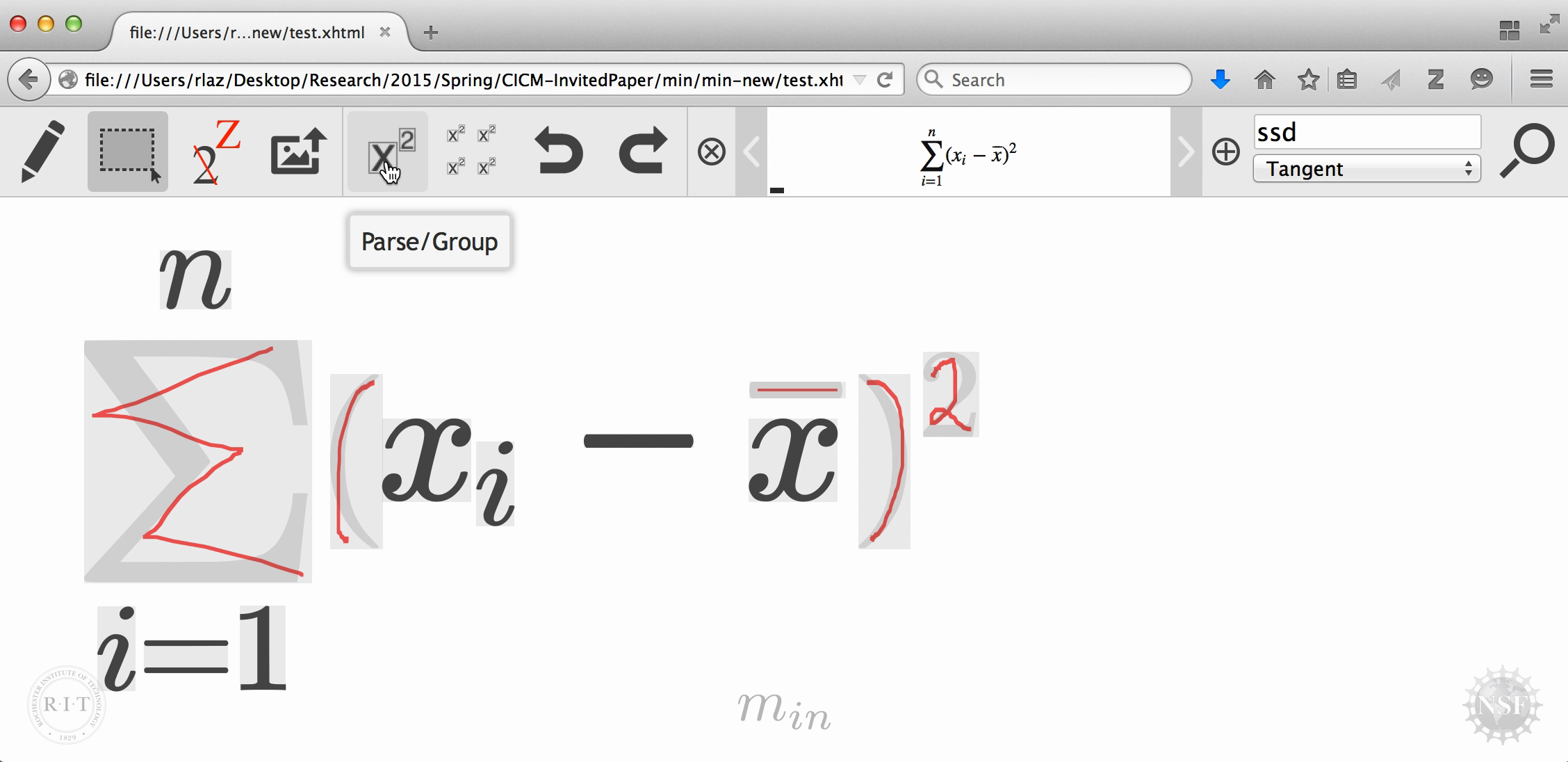}}\\
\caption{Combined Keyword (`ssd') and Formula Query in $m_{in}$ (Editing Mode). At top, from left-to-right
buttons are provided for symbol entry, selection and correction,
image import, parsing/grouping symbols, creating an expression grid (matrix), and undo/redo.
The `deck' stores a list of entered formulae, which may be combined (see
Figure \ref{fig:deck}). On pressing the search button (the magnifying glass),  \LaTeX~for the formula shown in the `deck' is concatenated with keywords and sent to a selected search engine  }
\label{fig:min}
\end{figure}

Figure \ref{fig:min} shows the $m_{in}$ search interface  which runs in standard web browsers on desktop and tablet computers (e.g. iPads \cite{min2012}).  $m_{in}$ is implemented in
Javascript and HTML, with symbol recognition and parsing performed by external web services.\footnote{\label{foot:source} Source code: \url{http://www.cs.rit.edu/~dprl/Software.html}}  In Figure \ref{fig:min} we see a text box for keywords at top-right, while the large white canvas at bottom is used to enter formulae. A list of formula are stored in the `deck,' the wide rectangular panel at the top of the interface. The deck has operations to add, remove, and switch between formulae.

$m_{in}$'s design seeks to allow mathematical non-experts to easily `draw' math expressions for queries by switching seamlessly between typing, freehand drawing, and importing formula images. Another design goal is providing clear, intuitive feedback for the recognized structure of an expression (i.e. the symbol layout tree). Our design was influenced by earlier math editing and recognition prototypes, particularly the pen-based Freehand Formula Entry System (FFES \cite{smithies99a,zanibbi2001aiding}), the vector graphics-based XPress system \cite{pollanen2007}, and the \emph{infty} math OCR system \cite{suzuki2004}. 

Figure 7 shows the entry of the formula in Figure \ref{fig:min}. A combination of typed \LaTeX~(e.g.  `$x_i - x$' in the top-left panel) and handwriting (shown as red lines) are used in queries. As handwritten symbols are recognized, they gradually fade and are replaced by the recognized symbol. \LaTeX~strings are replaced by MathJax renderings.  In the final step the symbol layout is parsed, and symbols are gradually moved in an animation to ideal positions. The fonts and locations for recognized symbols are obtained from Support Vector Graphics (SVG) MathJax output produced using the \LaTeX~string for recognized symbol layout. Handwriting is visible in the Editing mode, where pen/touch strokes appear above recognition results (see Figure \ref{fig:min}).

Figures \ref{fig:deck} and \ref{fig:matrix1} illustrate additional operations for image input, using the deck to store and combine formulae, matrix entry, and correcting symbol recognition errors.

\begin{figure}
\centering
\scalebox{1.35}{\includegraphics{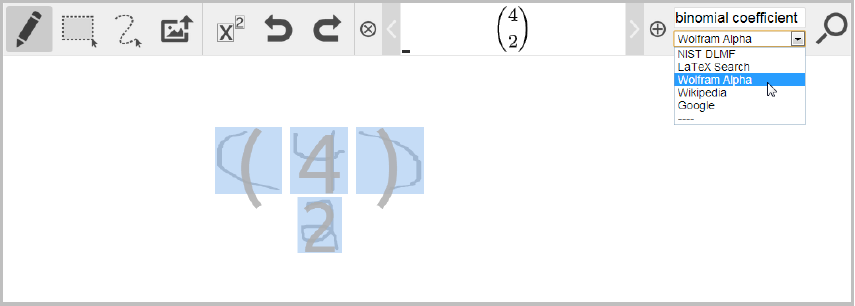}}
\caption{$m_{in}$ circa Spring 2013 (Drawing Mode \cite{Wangari2014}). The drop-down list of search engines is visible. The third button from the left at top corrects stroke groupings, and was later replaced by the symbol correction button.  Browser fonts drawn above handwritten strokes and simple symbol repositioning visualize recognition. In the new $m_{in}$ strokes gradually fade and are replaced by recognized symbols in draw mode  }
\label{fig:oldmin}
\end{figure}


\begin{figure}
\scriptsize
\begin{center}

\begin{tabular}{l l l}
\scalebox{0.05}{\includegraphics{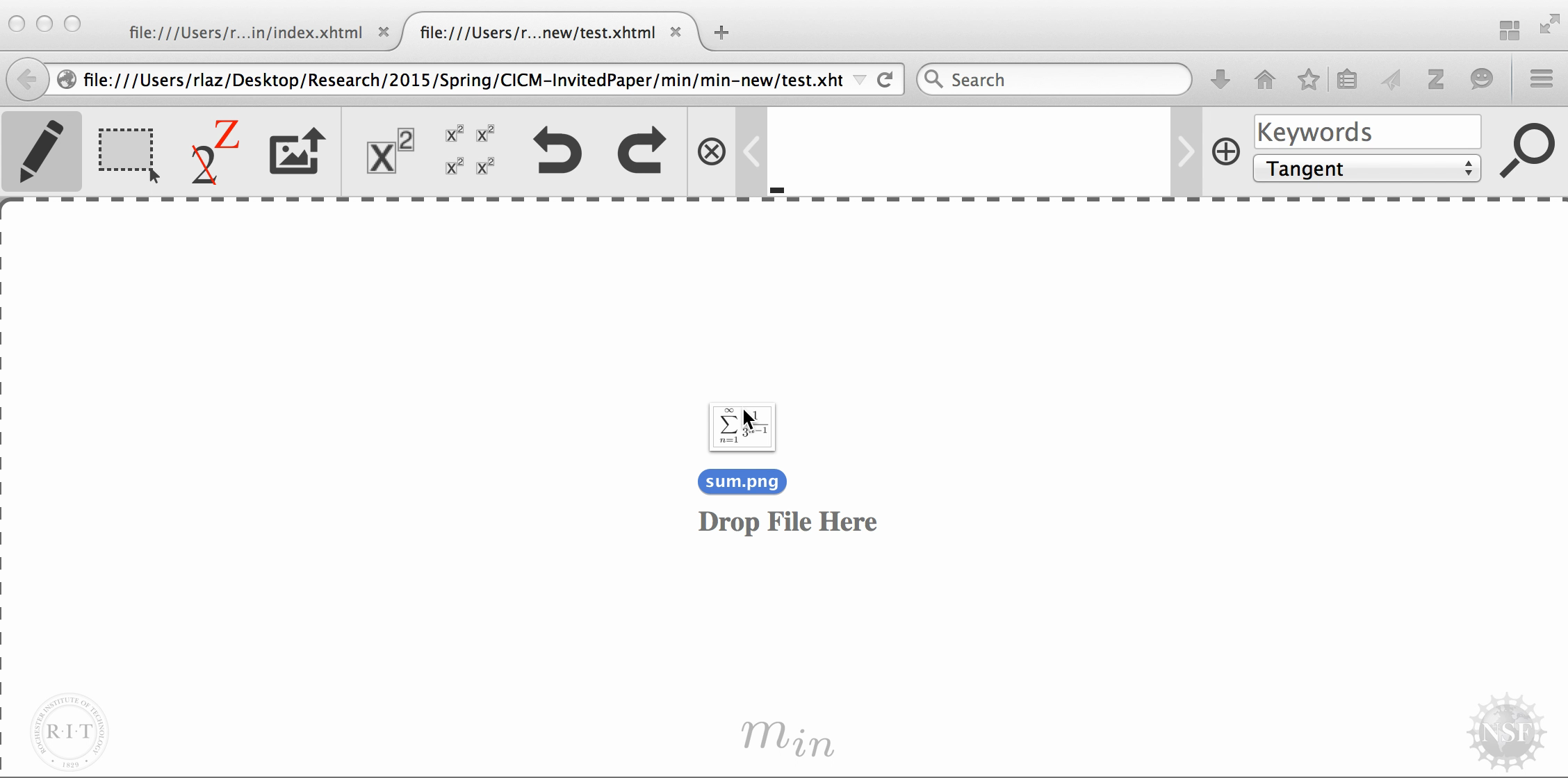}} &
\scalebox{0.05}{\includegraphics{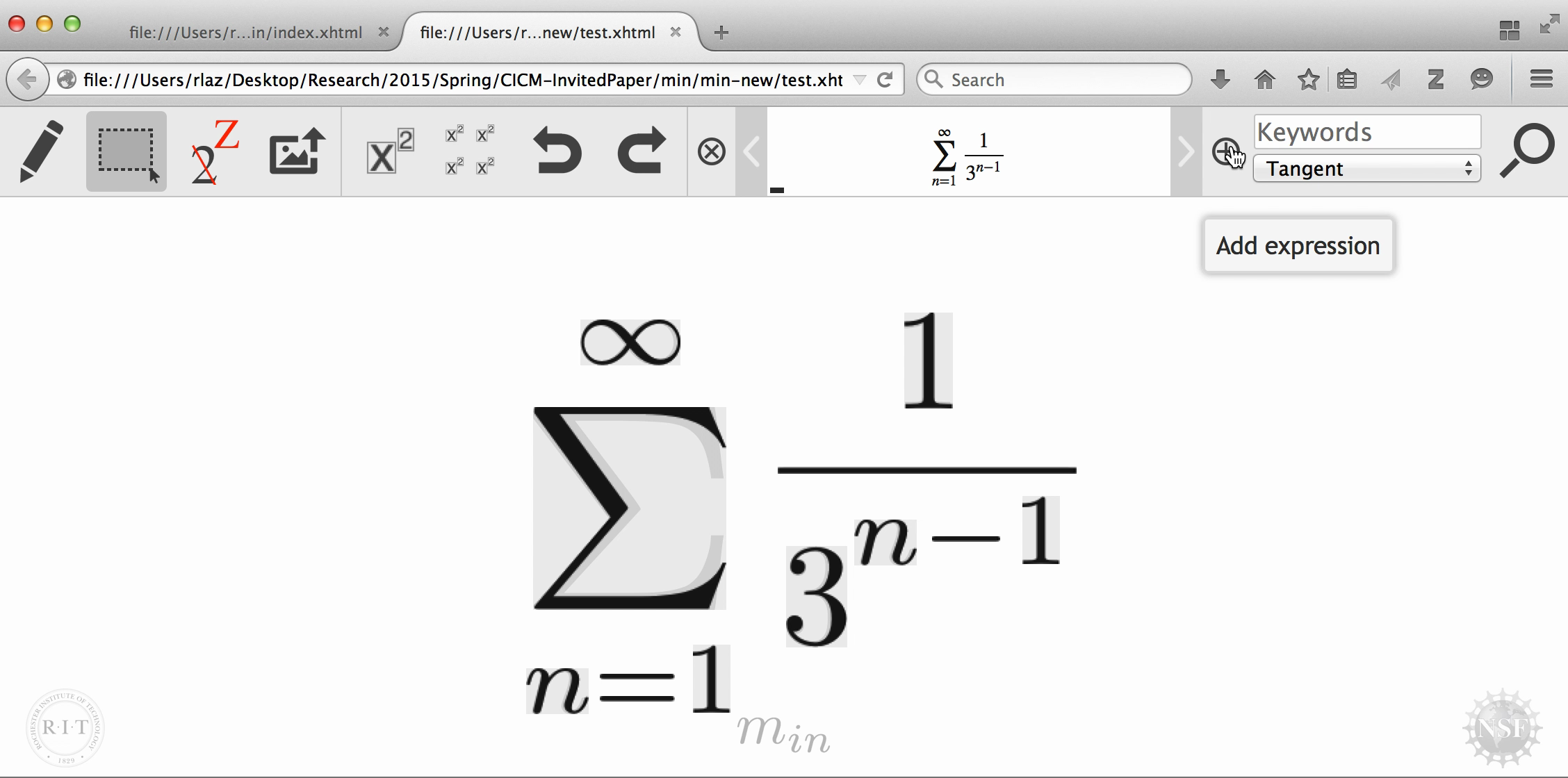}} &
\scalebox{0.05}{\includegraphics{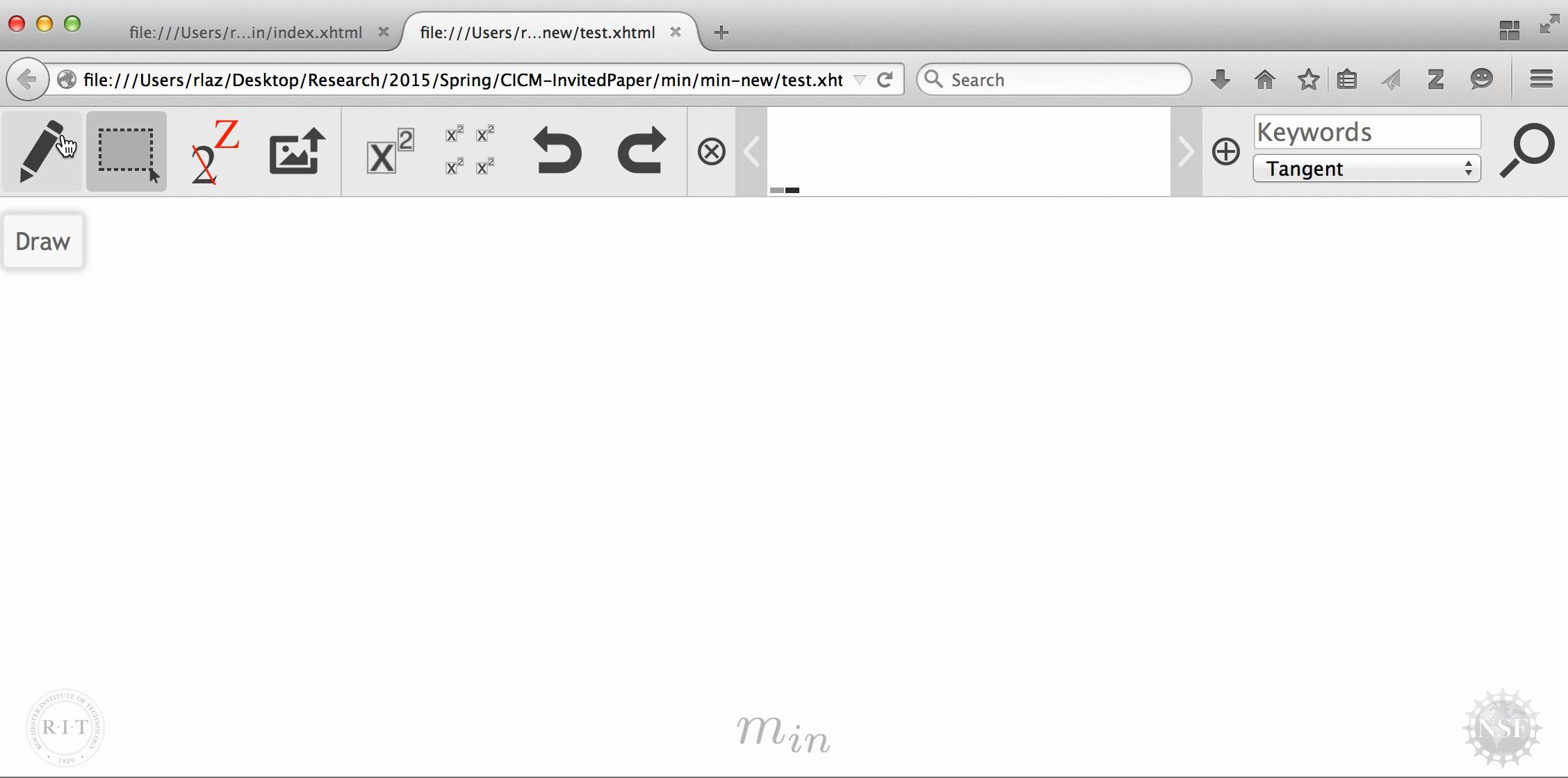}}
\\
a) Drag and drop image  &
b) Recognition result &
c) New deck panel; clear canvas \\
~\\
\scalebox{0.05}{\includegraphics{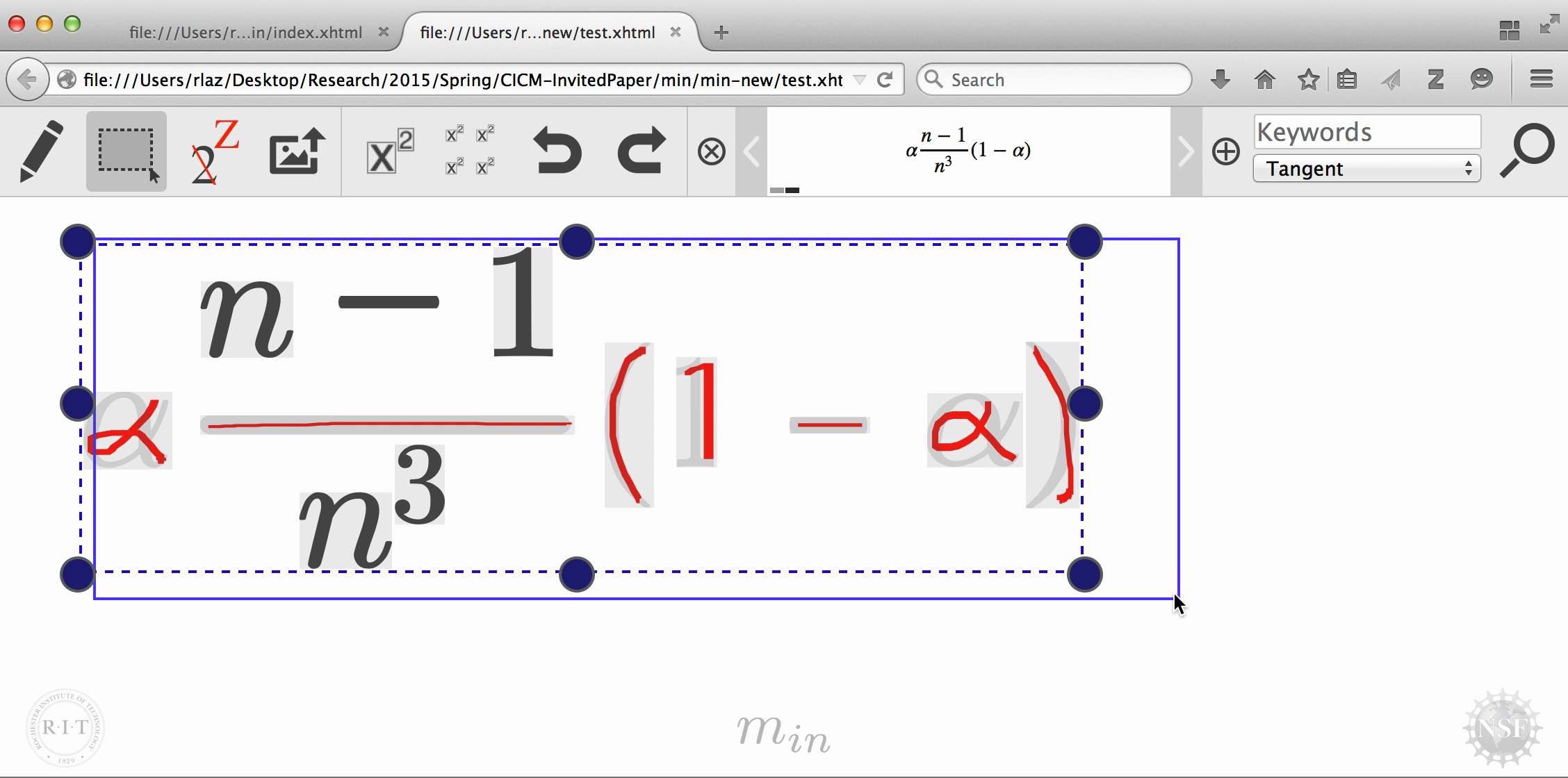}} &
\scalebox{0.05}{\includegraphics{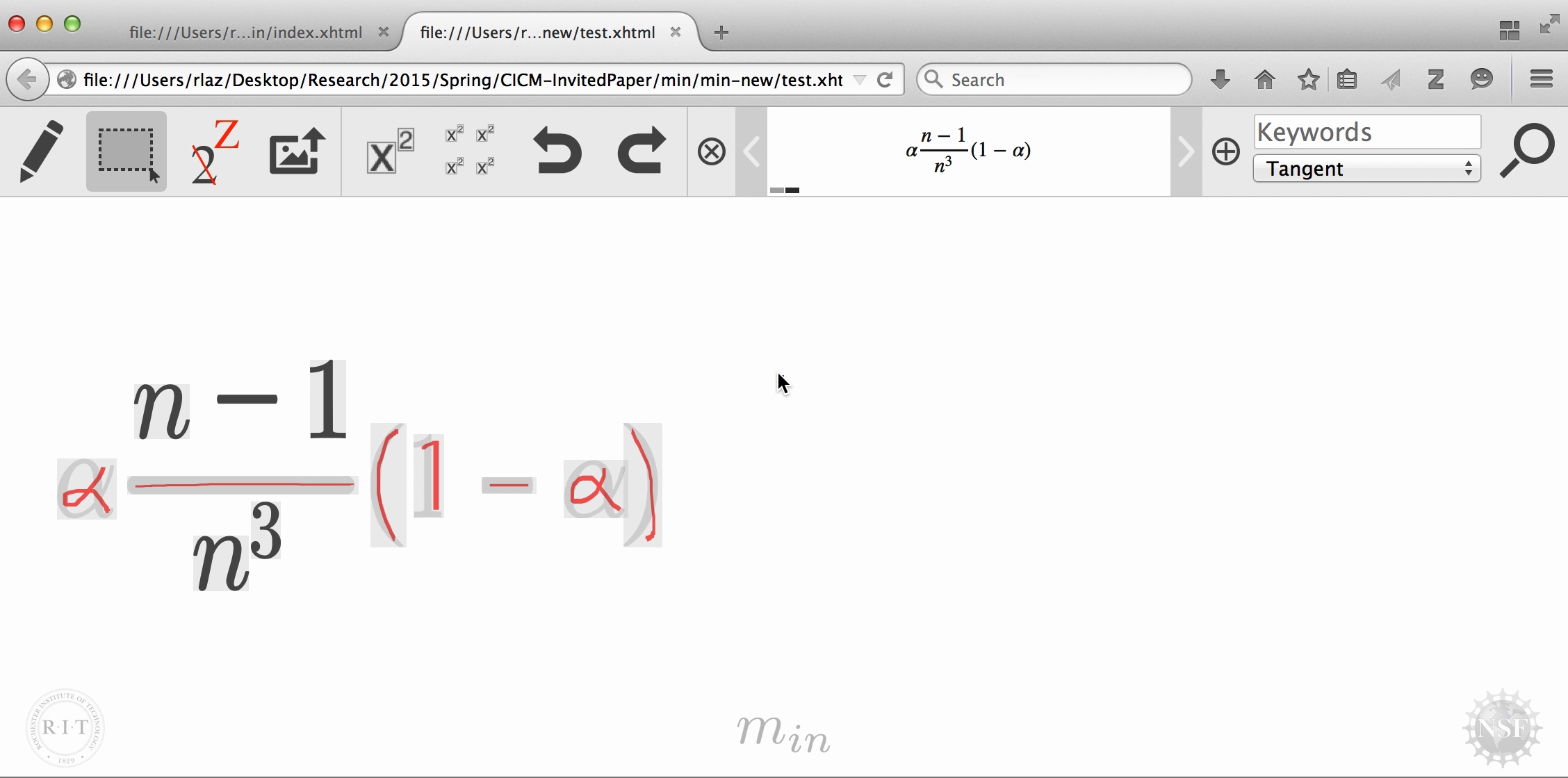}} &
\scalebox{0.05}{\includegraphics{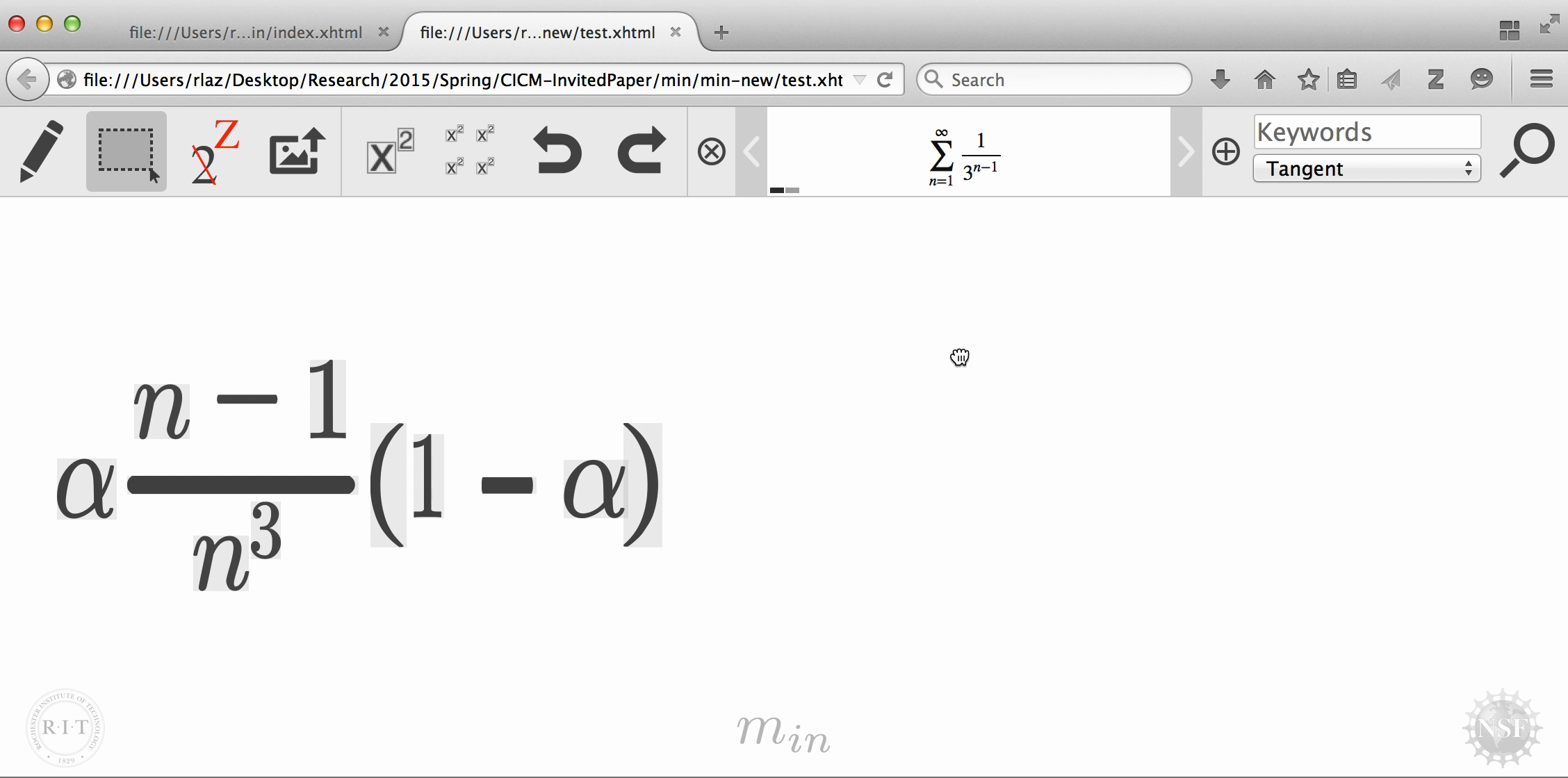}} 
\\
d) New expression &
e) Moved and resized &
f)  Drag from deck to canvas\\
~\\
\scalebox{0.05}{\includegraphics{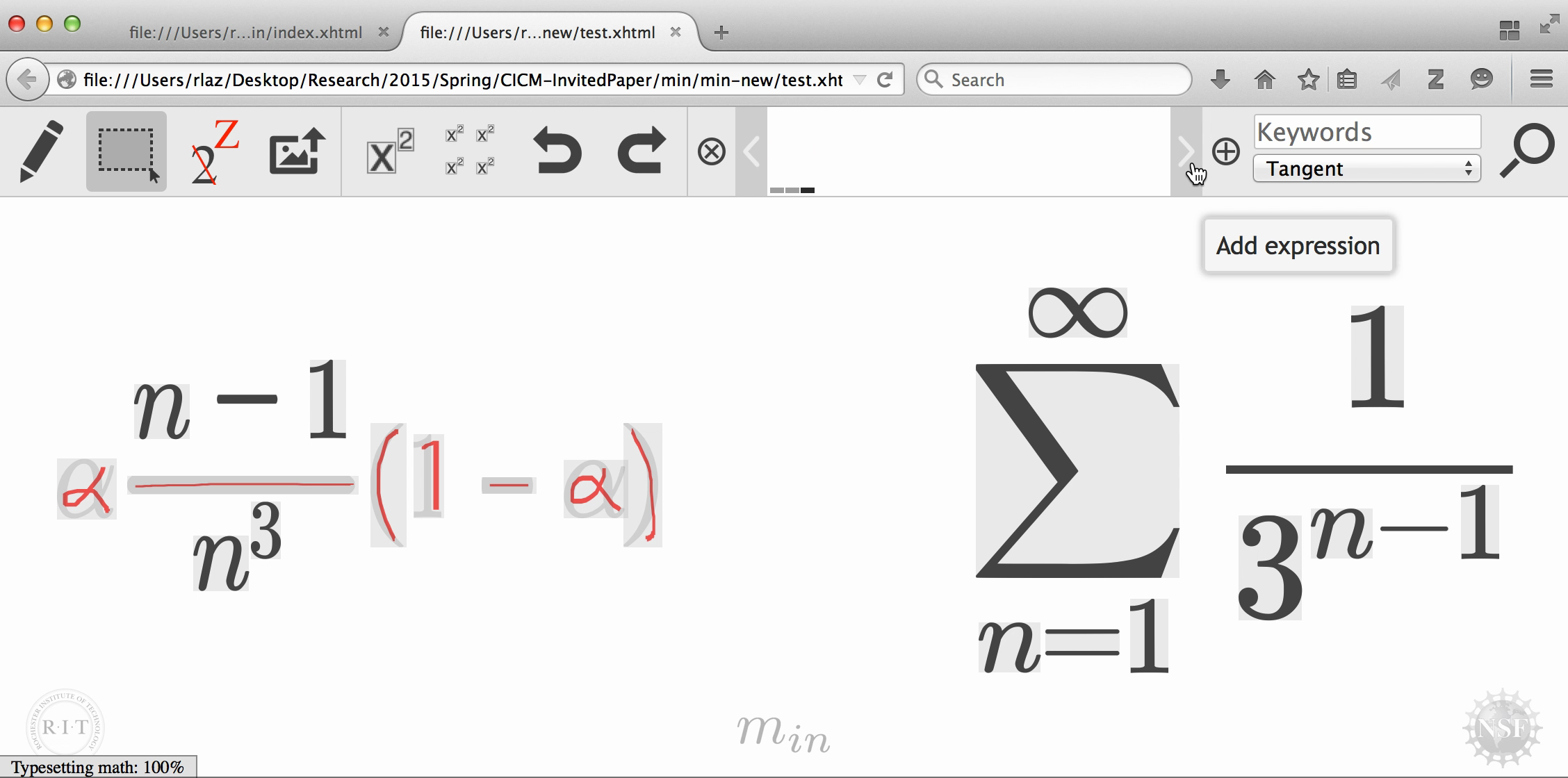}} &
\scalebox{0.05}{\includegraphics{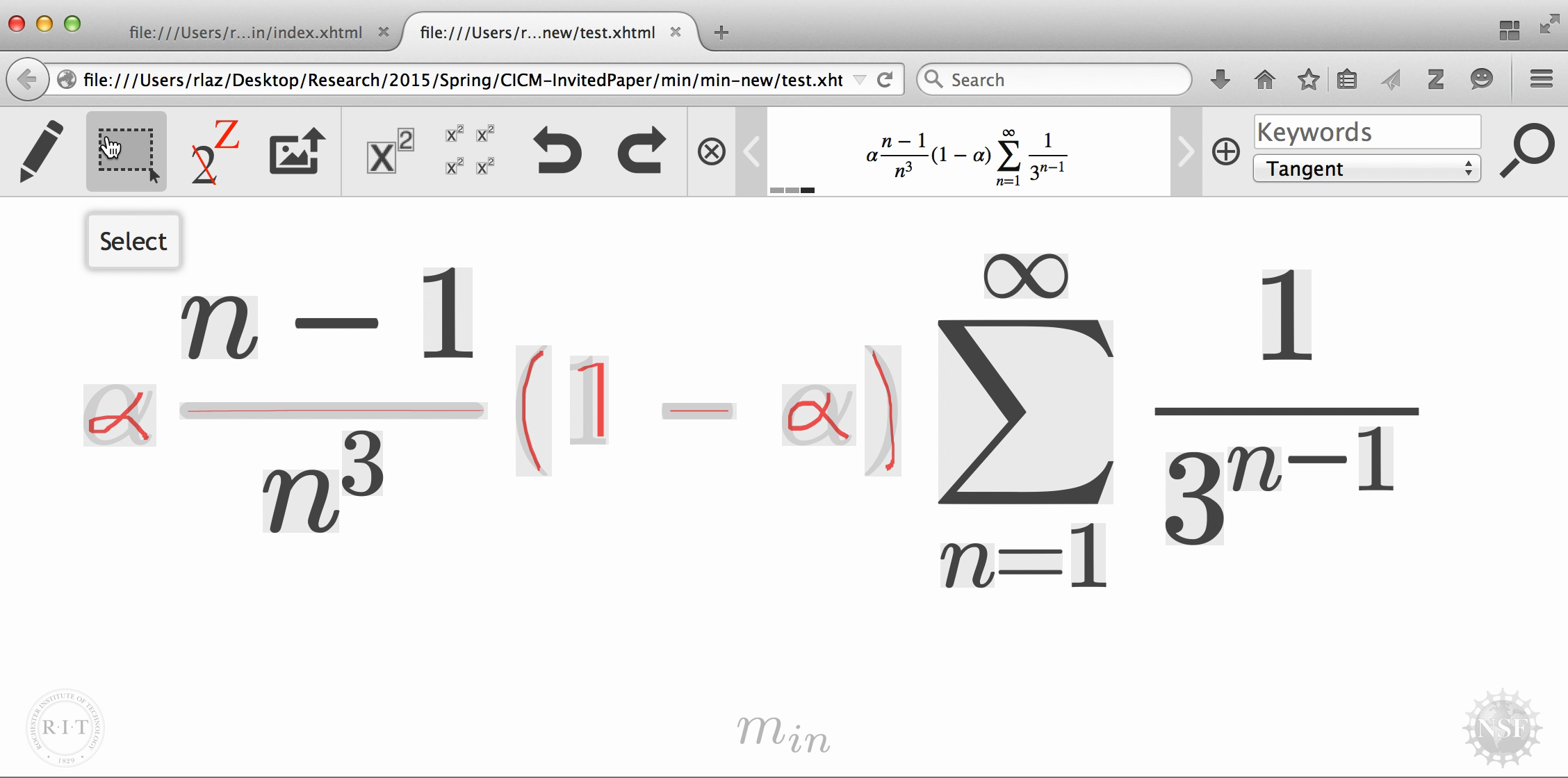}} &
\scalebox{0.05}{\includegraphics{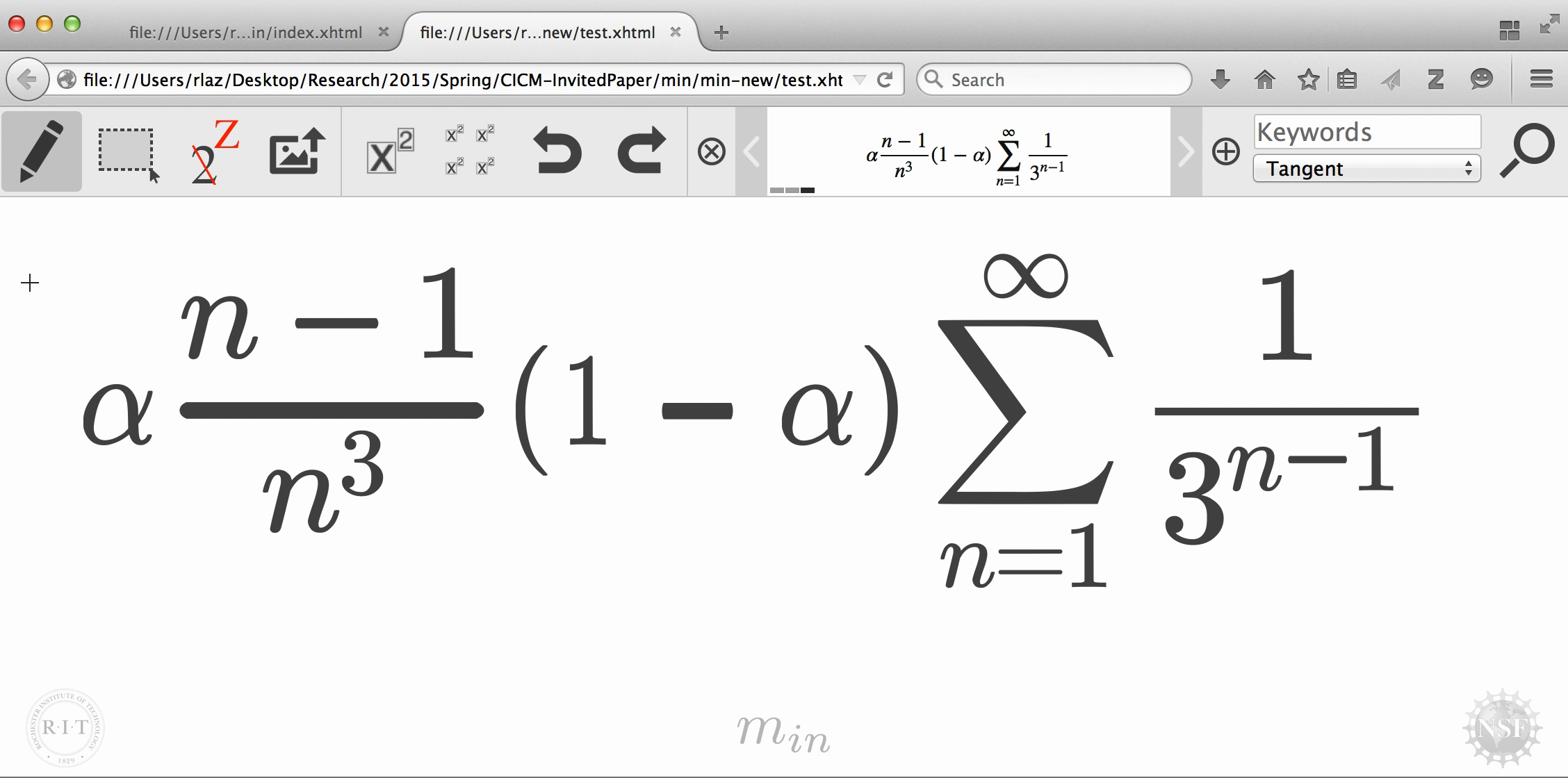}}
\\

g) After drop, add new panel &
h) Recognition result &
i) Draw view  
\\

\end{tabular}
\end{center}

\caption{Image Input and the Formula `Deck.' The panel (deck) showing images
at top of the interface stores formulae. An image creates
the first expression (a,b), which is then added to a second expression (c,d,e)
by dragging its panel from the deck to the canvas (f,g), and then storing
the combined result in a third slider panel (g,h,i)}
\label{fig:deck}
\end{figure}

\begin{figure}

\begin{center}
\scriptsize
\begin{tabular}{l l l}
\scalebox{0.05}{\includegraphics{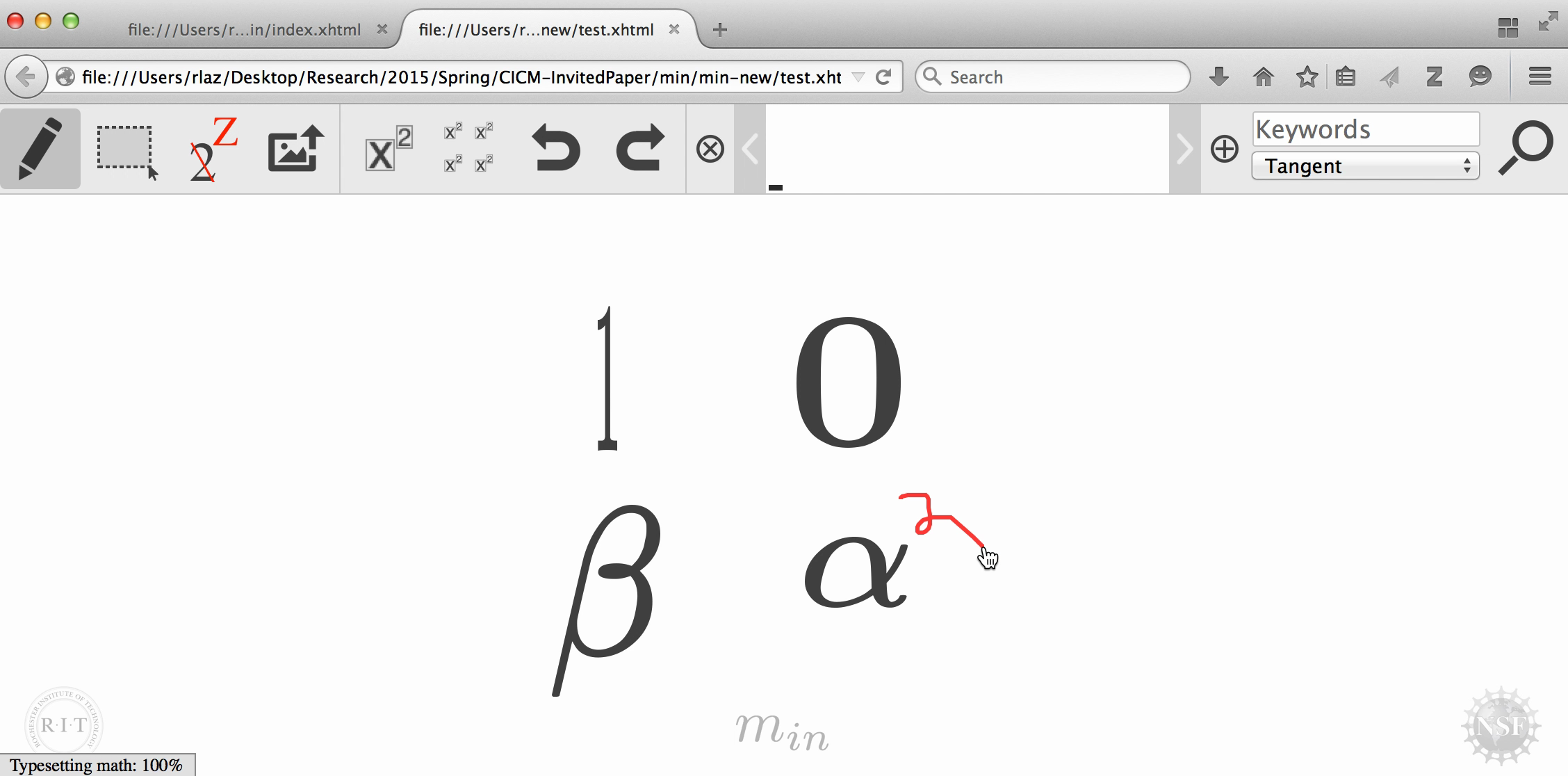}} &
\scalebox{0.05}{\includegraphics{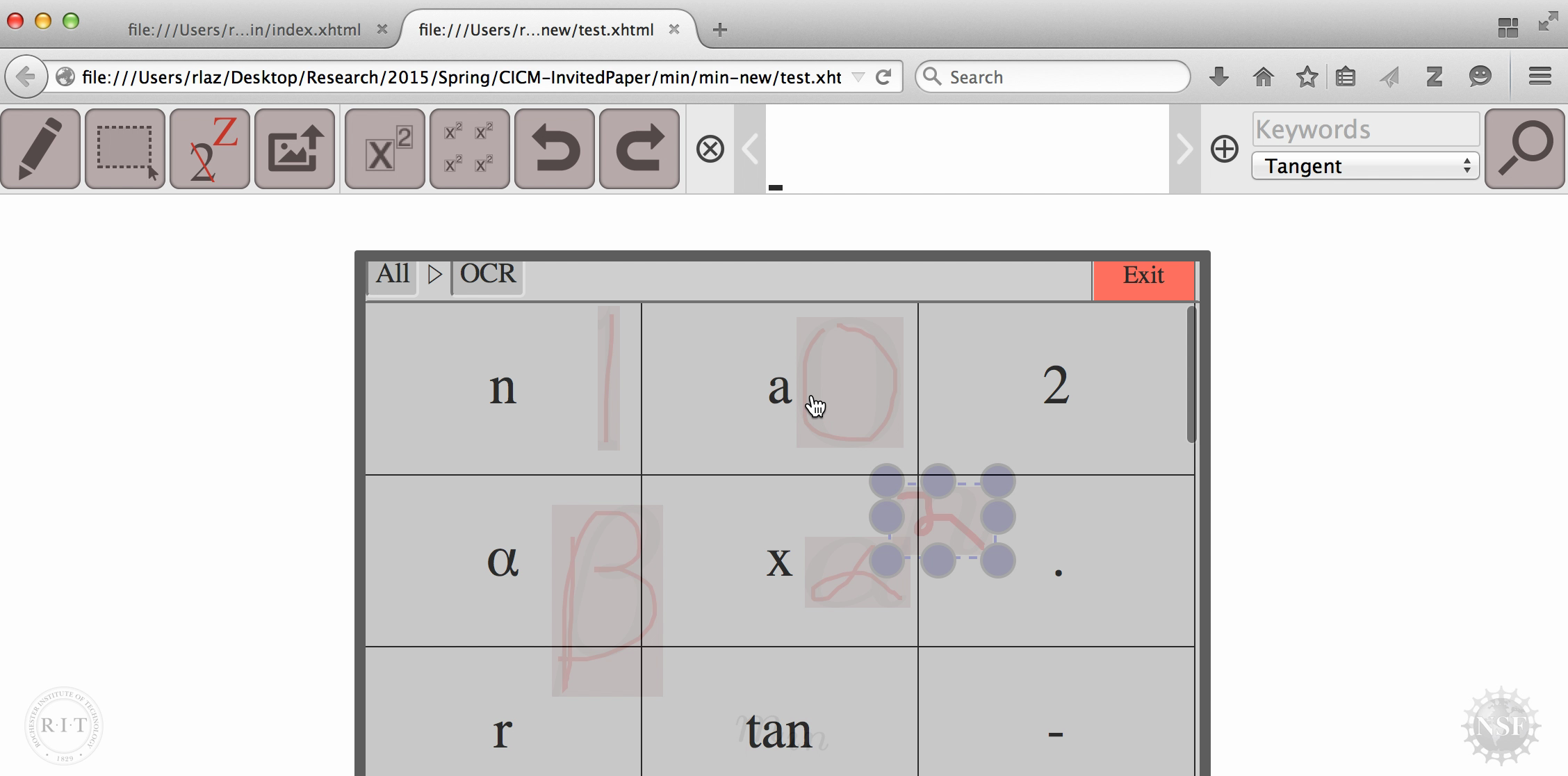}} &
\scalebox{0.05}{\includegraphics{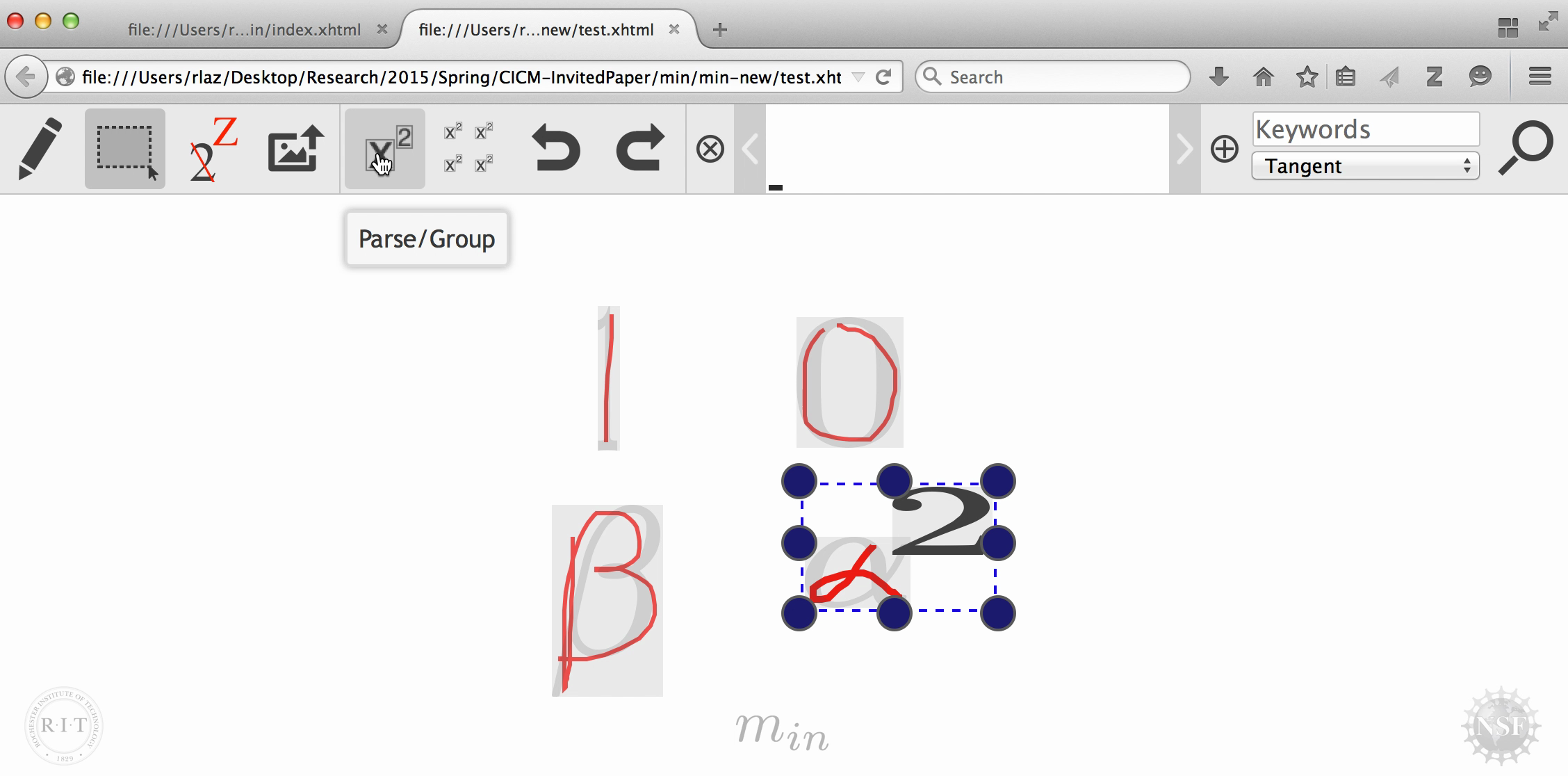}}\\
a) Drawing a `2' &
b) Correct misrecognition as `n' &
c) Select subexpression
\\
~\\

\scalebox{0.05}{\includegraphics{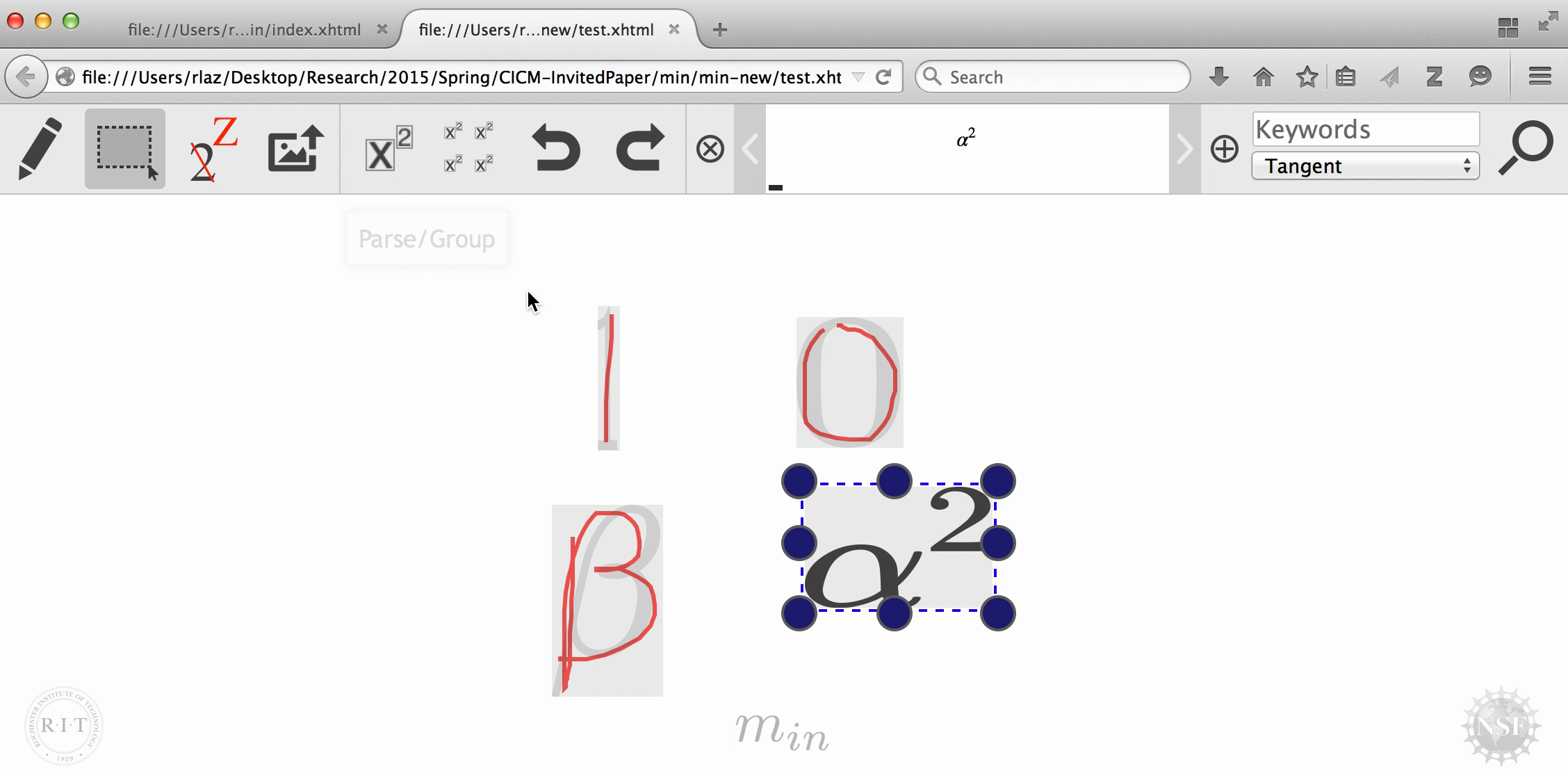}} &
\scalebox{0.05}{\includegraphics{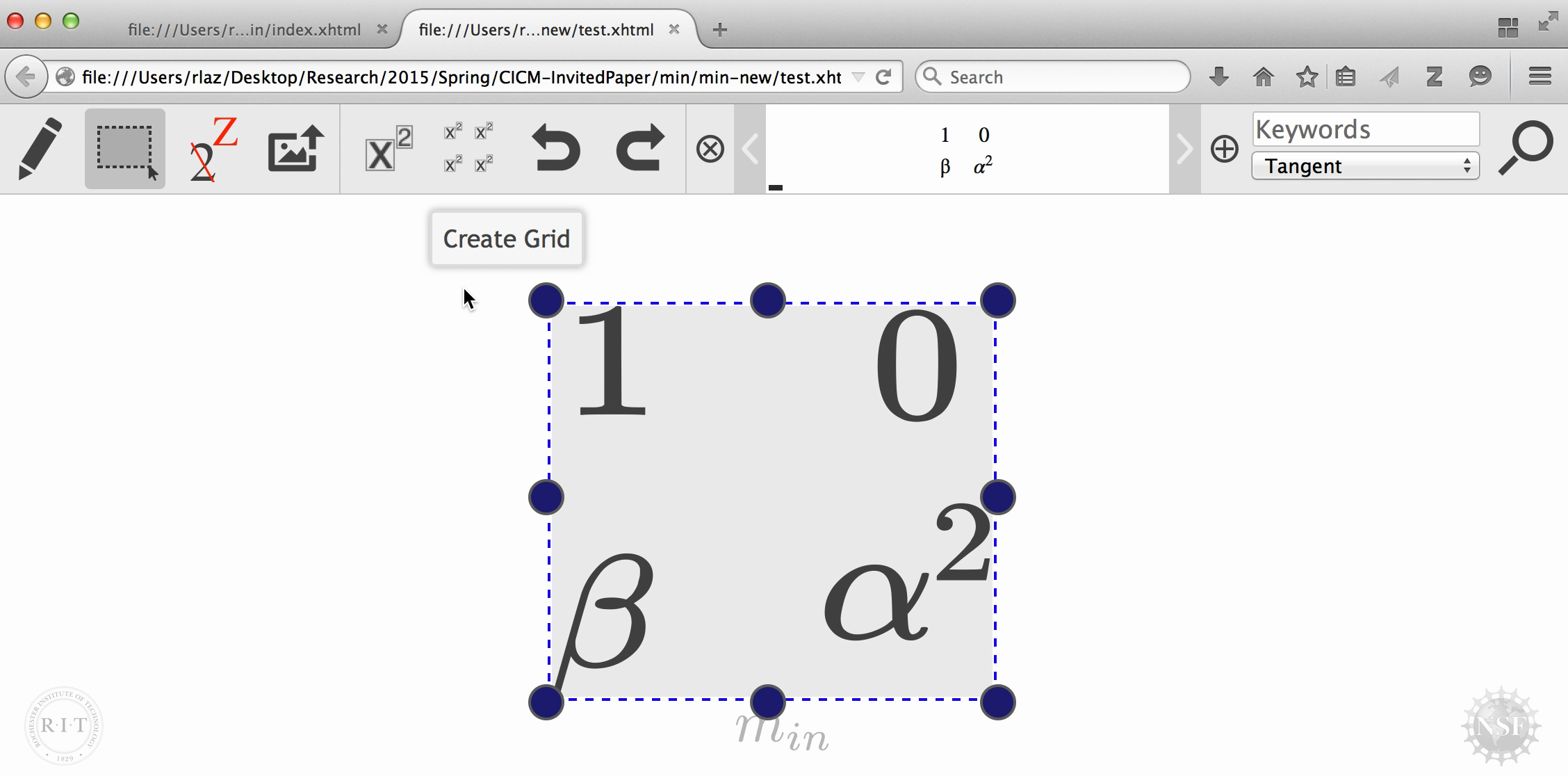}} &
\scalebox{0.05}{\includegraphics{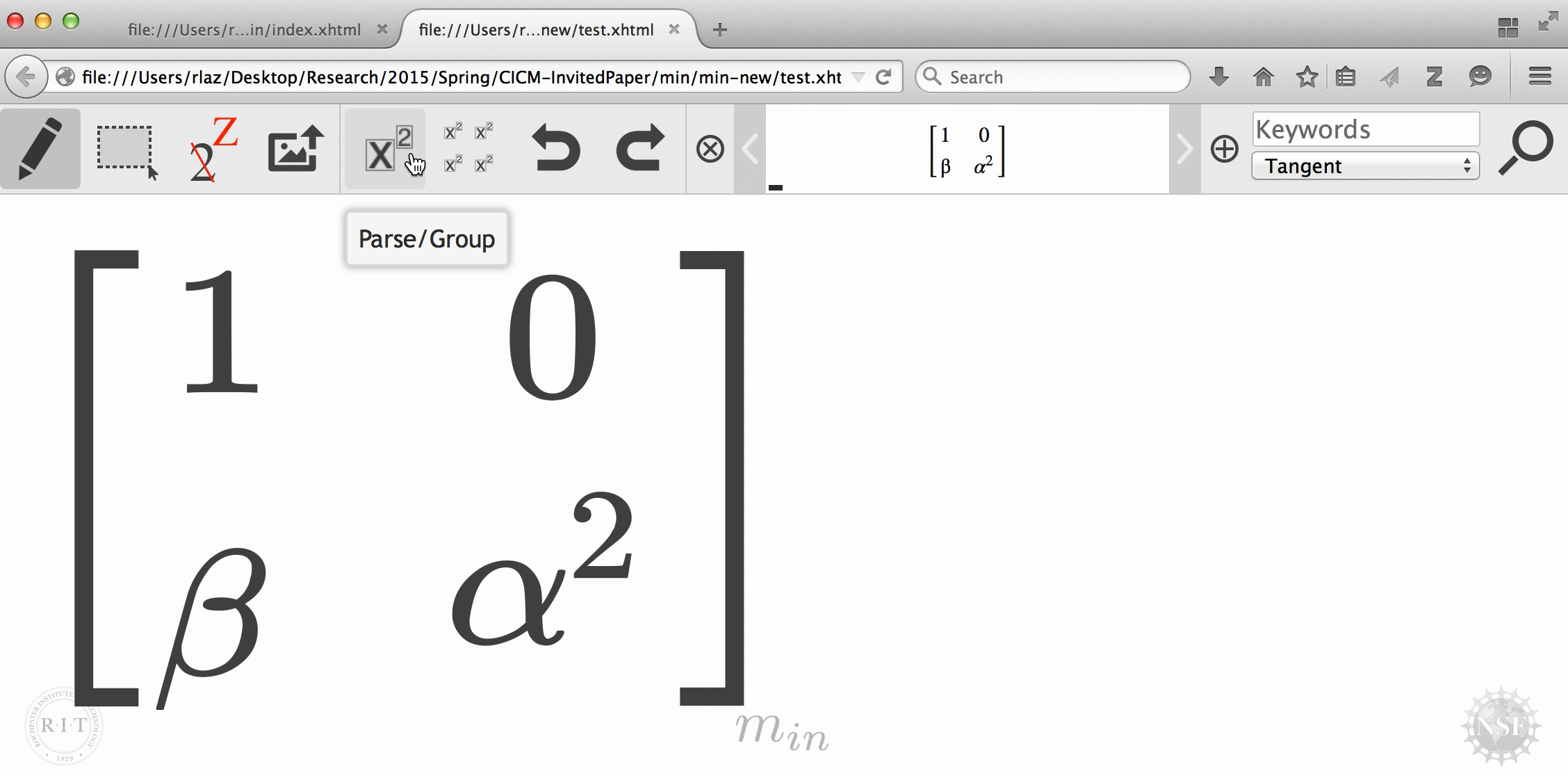}}
\\
d) Parsed subexpression &
e) Create formula grid (matrix) &
f)  Draw braces and parse\\
\end{tabular}

\end{center}

\caption{Matrix Entry and Symbol Correction. As shown in this
example, $m_{in}$ allows subexpressions to be grouped separately (c,d) or as a grid of expressions
(e).  Symbol recognition errors are corrected using a 
transparent
pop-up window (b). The window appears after selecting a symbol and then
pressing the `relabel' button at top 
}
\label{fig:matrix1}
\end{figure}

\subsection{Human Studies: Formula Entry Operations and Recognition Visualization}

A pair of human studies have influenced the design of $m_{in}$. The first study compared  visualizations of recognition results for handwritten formulae. In the first condition, results were shown separately from user input in a rendered \LaTeX~image, and in the second condition handwritten symbols were gradually rescaled and moved to ideal positions using a \emph{style-preserving morph}  \cite{zanibbi2001aiding}. Overall, participants found results from the rendered image clearer, but surprisingly there was no significant difference in entry time for users using the image or morphing feedback, despite symbol recognition results not being visible in the morphing condition unless individual symbols were selected. Also, in the image condition some participants became stuck, as they were unable to find where their expression was recognized incorrectly (this finding has been replicated in other studies since; see \cite{zanibbi2012recognition}). This did not happen when the participant's symbols were `morphed' in-place. 

In $m_{in}$ a style-preserving morph is performed when a button to recognize expression structure is pressed (the `Parse/Group' button). This `morph' is actually a modified version, described below.

The second human study evaluated an earlier version of $m_{in}$ (see Figure \ref{fig:oldmin}), and identified opportunities for improvement  \cite{Wangari2014}. In particular, the undergraduate college students that participated in the study found that while symbol recognition results were now visible when drawing (they would appear above user strokes, see Figure \ref{fig:oldmin}), this cluttered the canvas, making it difficult to see errors. The symbol placement from the original style preserving morph is also coarse and sometimes confusing (e.g. making adjacent symbols appear subscripted \cite{zanibbi2001aiding}). To address these issues, in the new $m_{in}$ recognized symbols replace handwritten strokes in the drawing view using a gradual fade, and the target positions for morphing are defined using rendered \LaTeX. To avoid loss of context and interfering with users' `mental maps,' handwritten strokes remain visible in the editing mode, with strokes shown in red above characters for recognized symbols (see Figure \ref{fig:min}).

Participants in the second study were also shown a tool for stroke grouping to correct symbol segmentation errors at the beginning of each session, but there was almost no use of this tool, with participants instead deleting and redrawing symbols if they were segmented incorrectly. Participants also had difficulty remembering that double-clicking/tapping on a symbol brings up the symbol correction menu (see Figure \ref{fig:matrix1}b).  As a result, the stroke grouping button was replaced by a button for relabeling selected symbols in the new version of $m_{in}$. 

In the future we hope to carry out additional studies to evaluate the new interface, and in particular the utility of the formula deck when working with multiple expressions, and the new matrix entry operations. We feel that we have made some progress, but questions about which formula editing and correction operations to include in the interface, and how best to visualize recognition results remain.

\subsection{$m_{in}$ System Architecture and Recognition Modules}

Figure \ref{fig:minarch} presents a global view of $m_{in}$'s architecture. Users input keywords as text, and math using a combination of \LaTeX, handwritten symbols and images. There are two primary data structures that define the interpretation of a formula on the canvas:  a list of symbols and their locations, and  the recognized symbol layout tree for the formula. For clarity, we do not show the formula `deck' in Figure \ref{fig:minarch} (see Figure 4 for an illustration of the deck). When the user clicks on the search button, keywords and the current expression shown in the deck are concatenated in a query string, which is then sent to the currently selected search engine.

\begin{landscape}
\begin{figure}
\begin{center}

\scalebox{0.4}{\includegraphics[clip=true,trim=0.5in 0in 0in 0in]{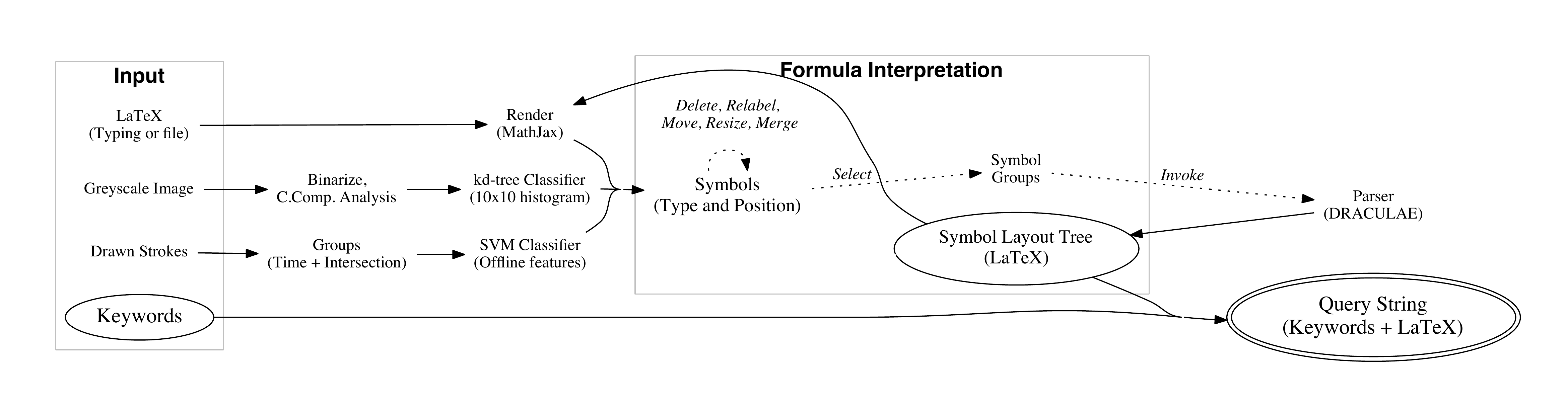}}
\end{center}
\vspace{-0.45in}
\caption{$m_{in}$ Architecture. Users interact with the symbol through loading images, drawing symbols, and typing. User interactions are shown using dotted arrows, and solid arrows
represent automated processing.} 

\label{fig:minarch}
\end{figure}

\begin{figure}[!h]
\begin{center}

\begin{tabular}{l l l l l l}

\scalebox{0.04}{\includegraphics{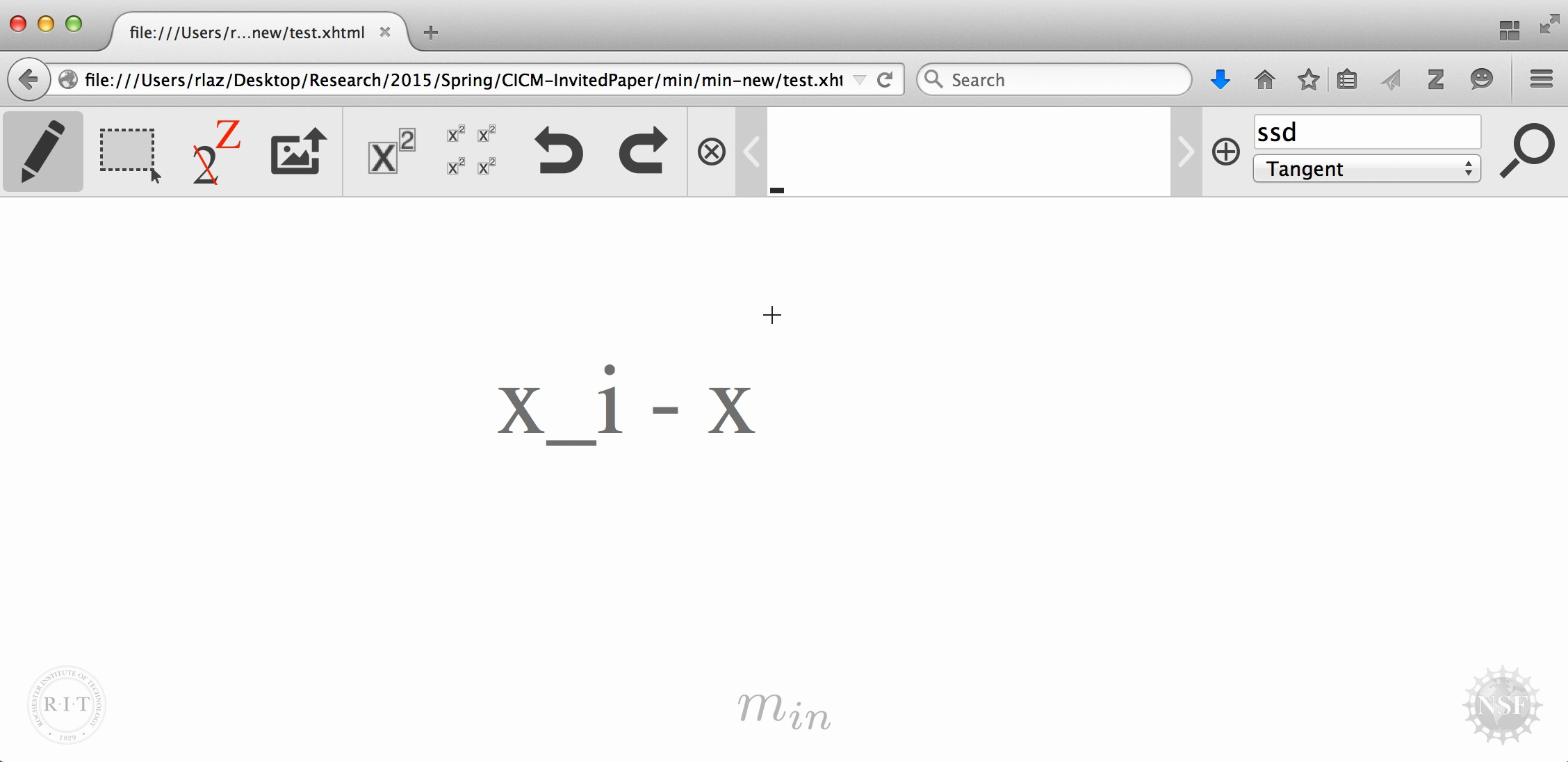}} &
\scalebox{0.04}{\includegraphics{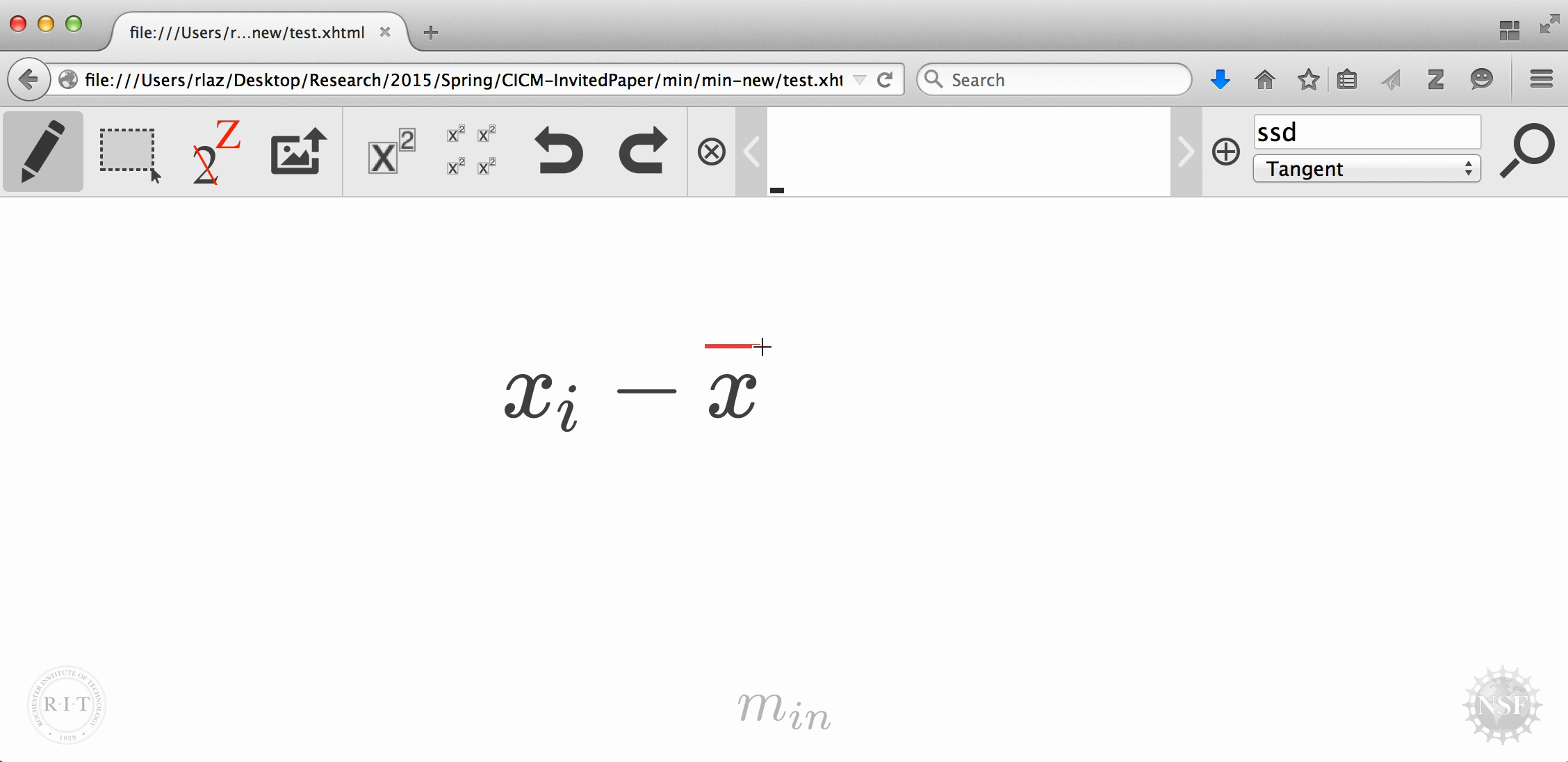}} &
\scalebox{0.04}{\includegraphics{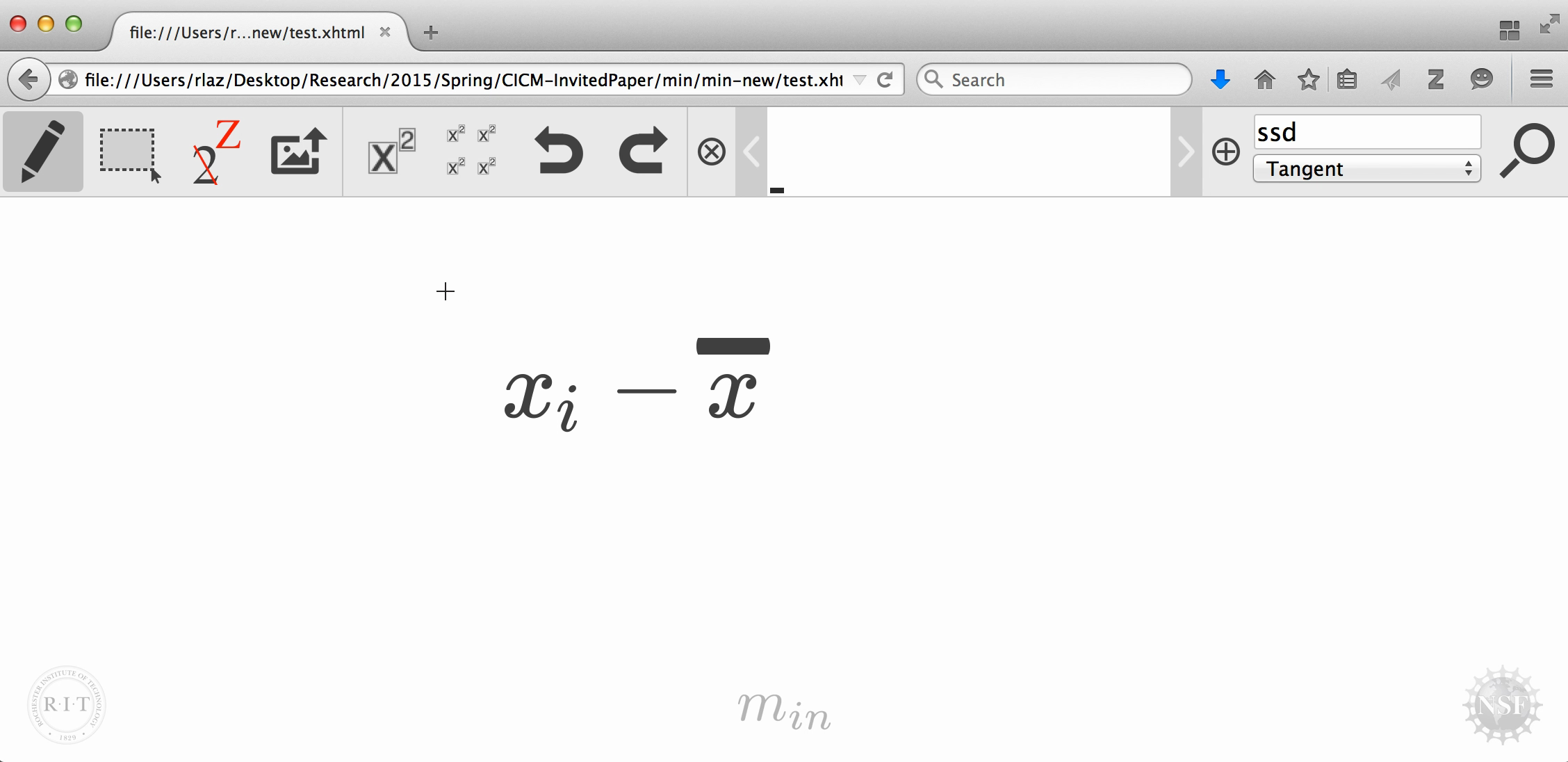}} &

\scalebox{0.04}{\includegraphics{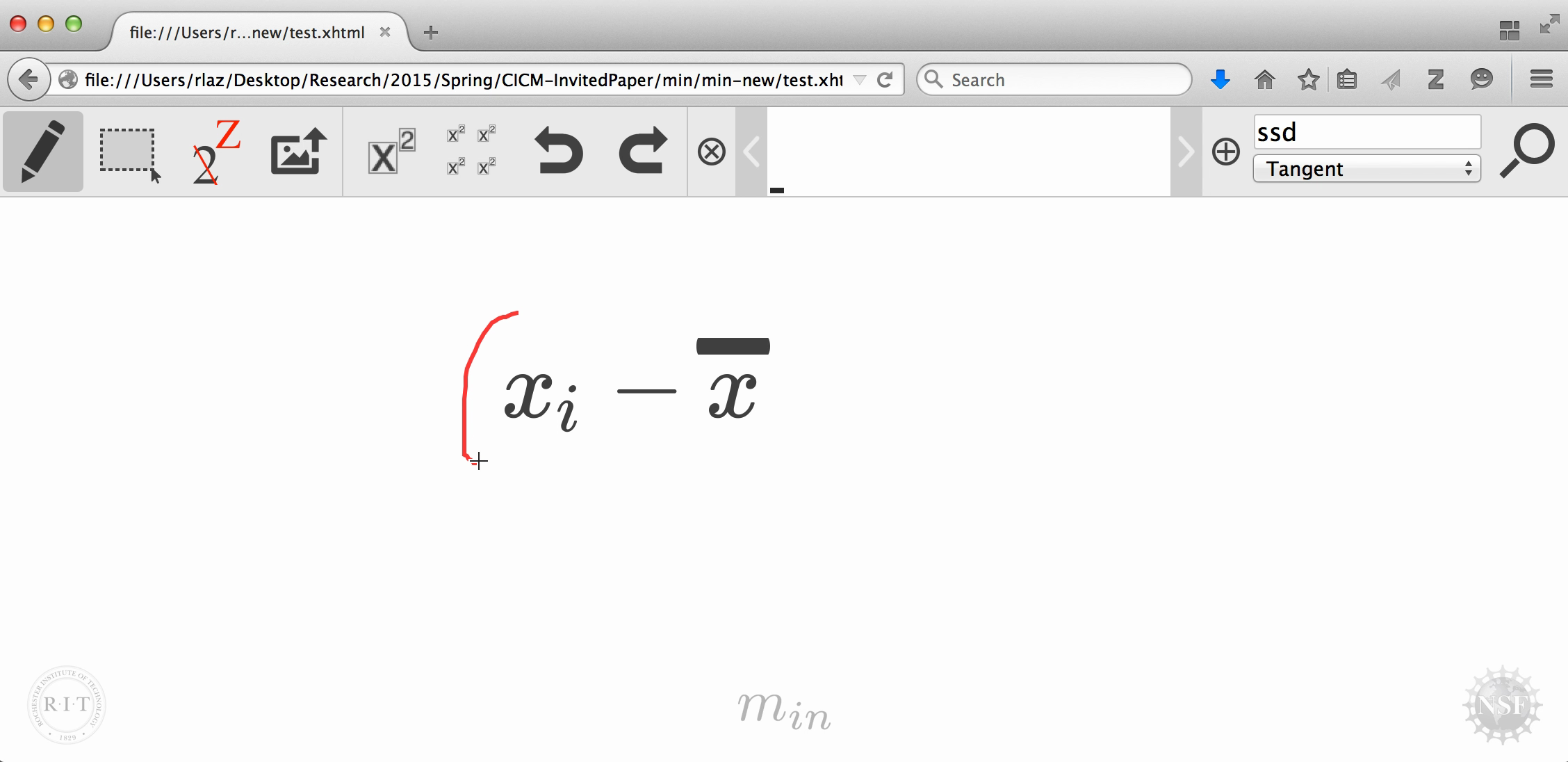}} &
\scalebox{0.04}{\includegraphics{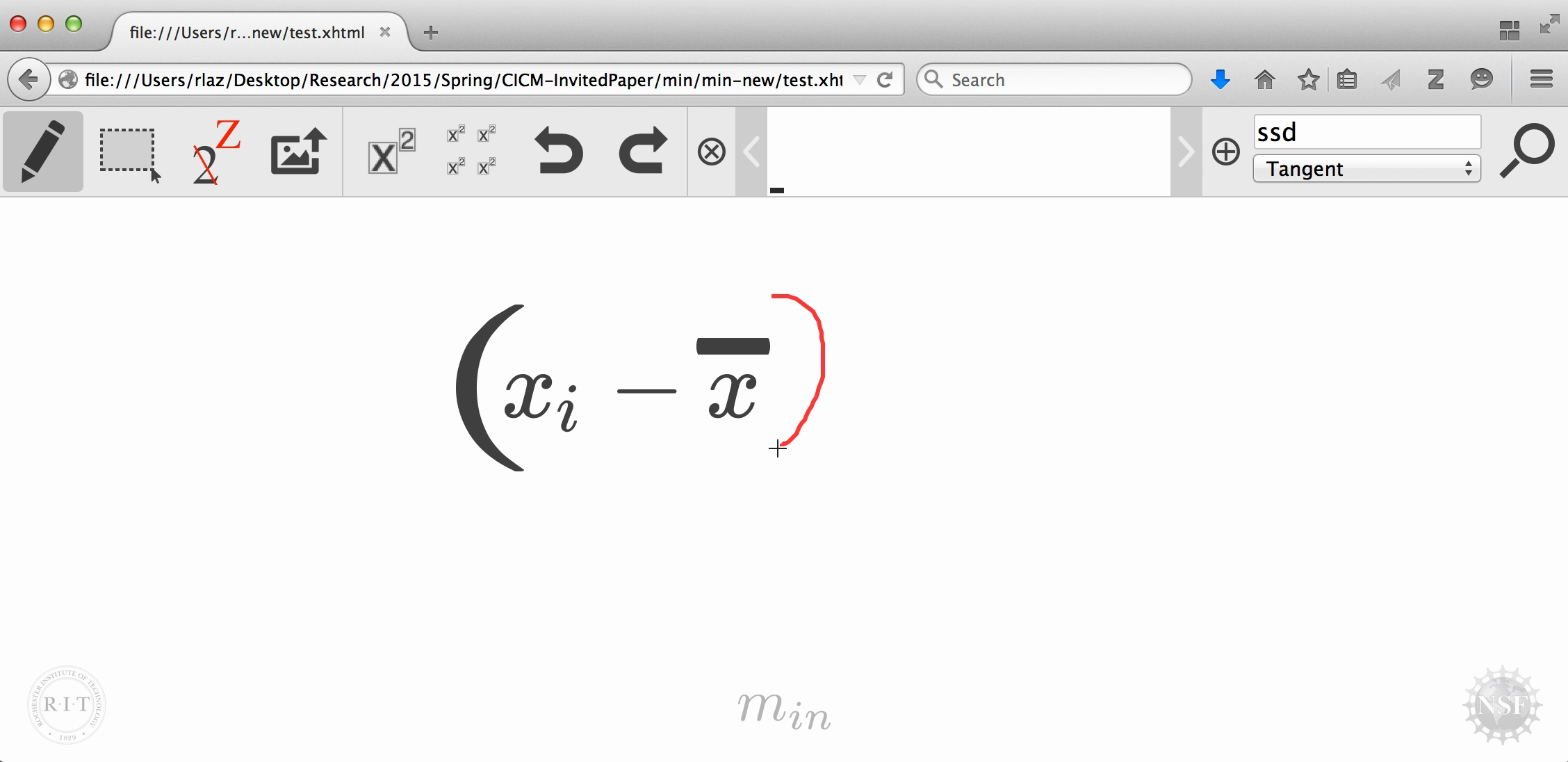}} &
\scalebox{0.04}{\includegraphics{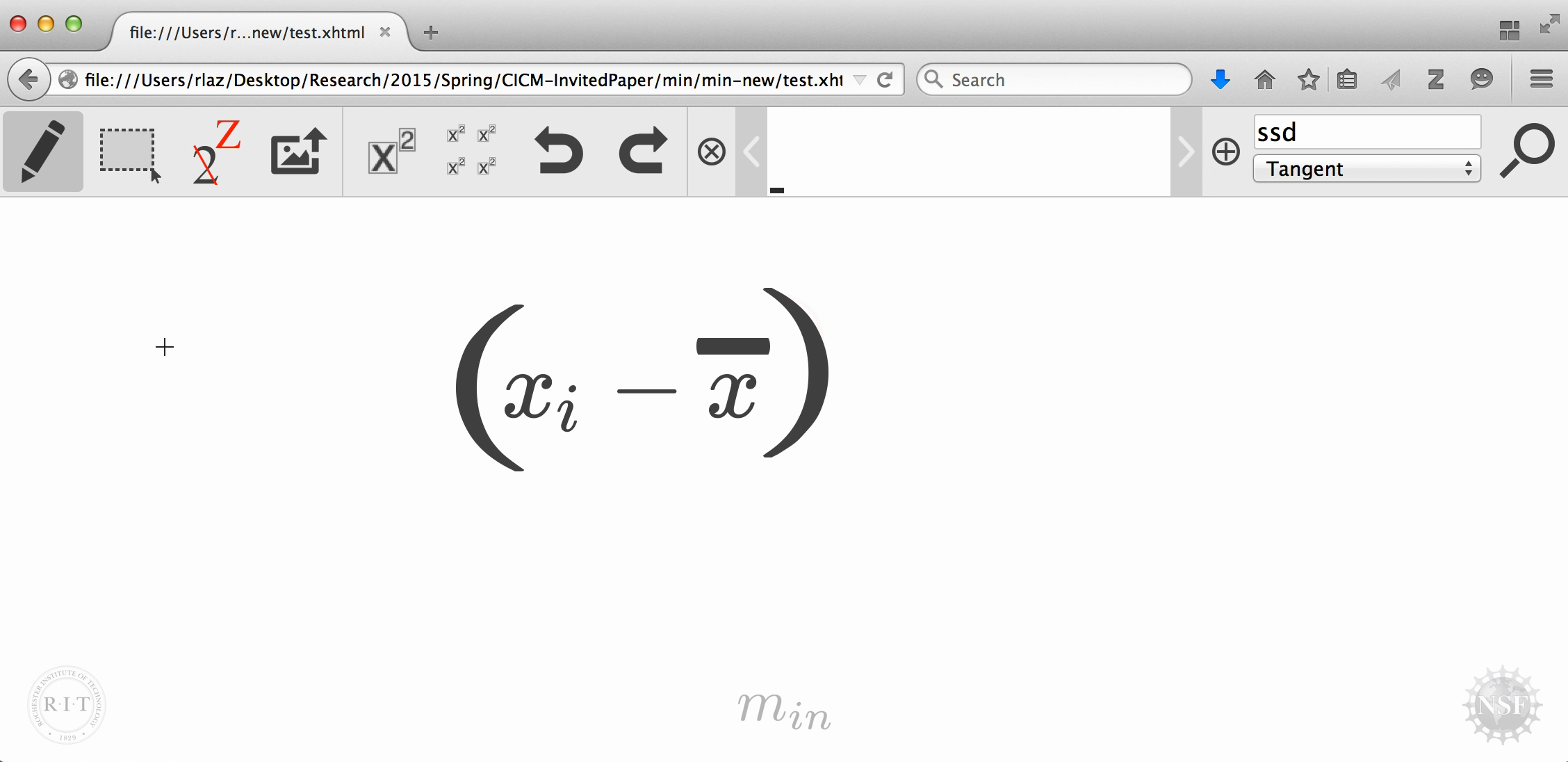}} \\

\scalebox{0.04}{\includegraphics{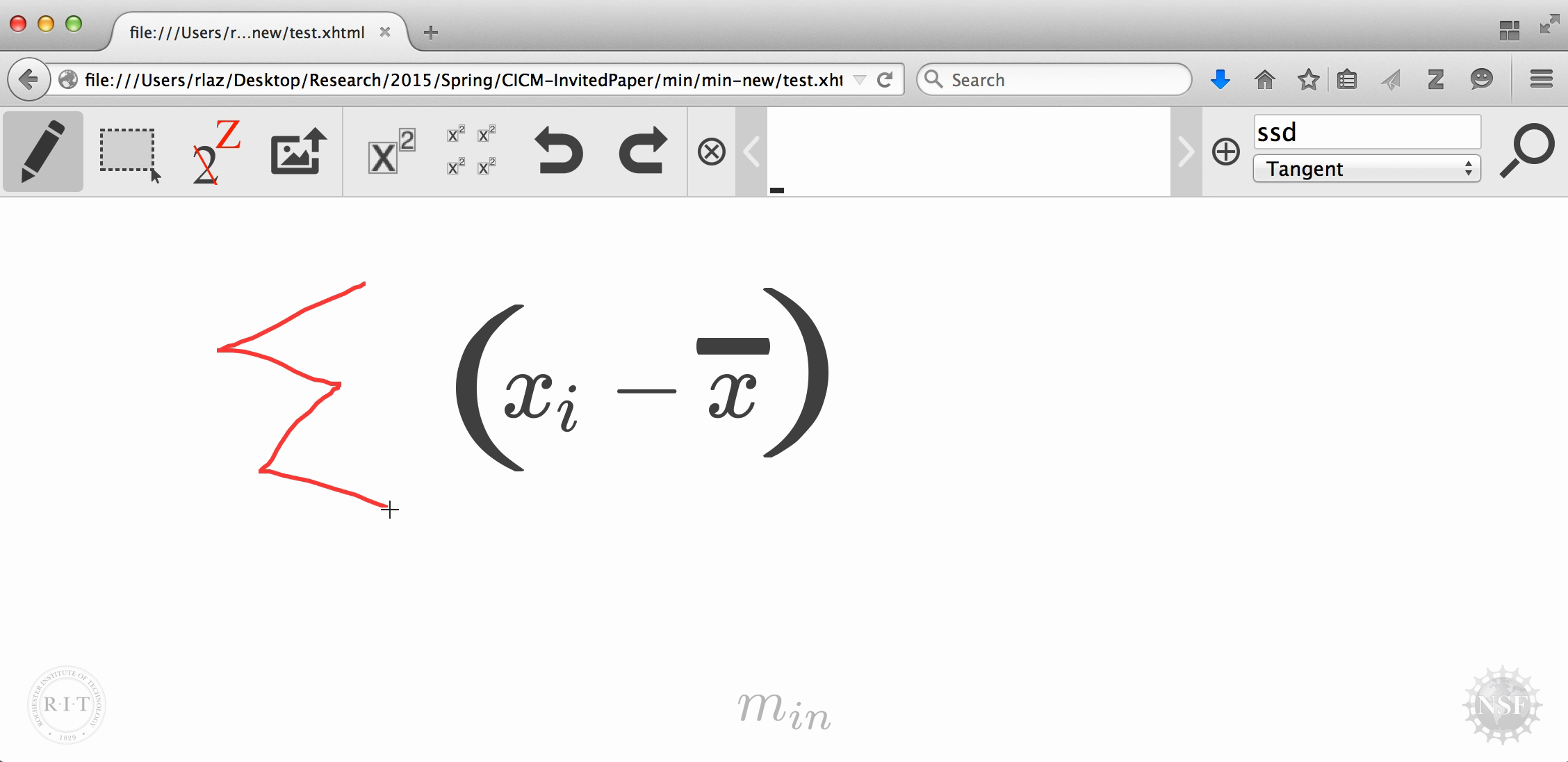}} &
\scalebox{0.04}{\includegraphics{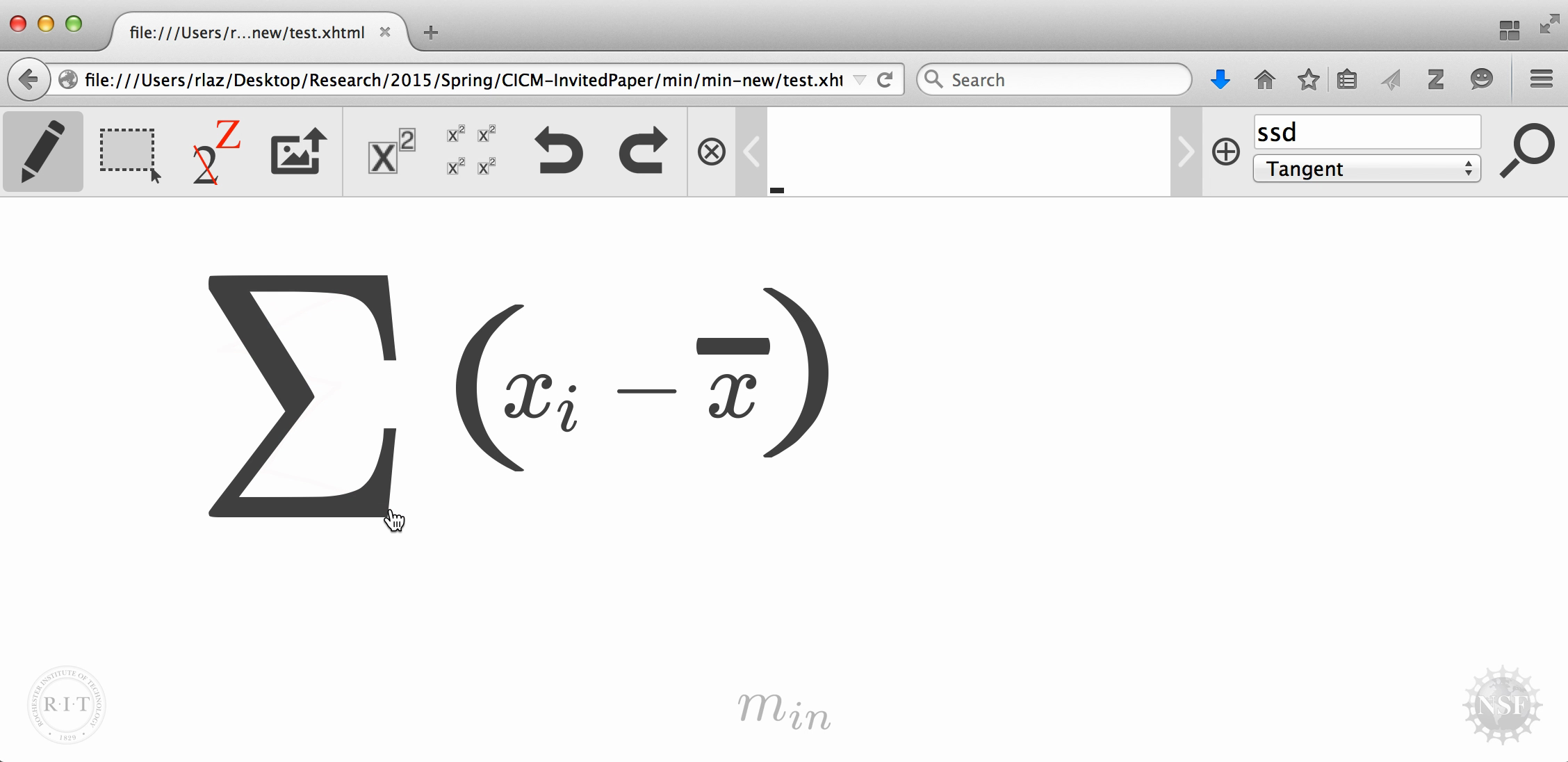}} &
\scalebox{0.04}{\includegraphics{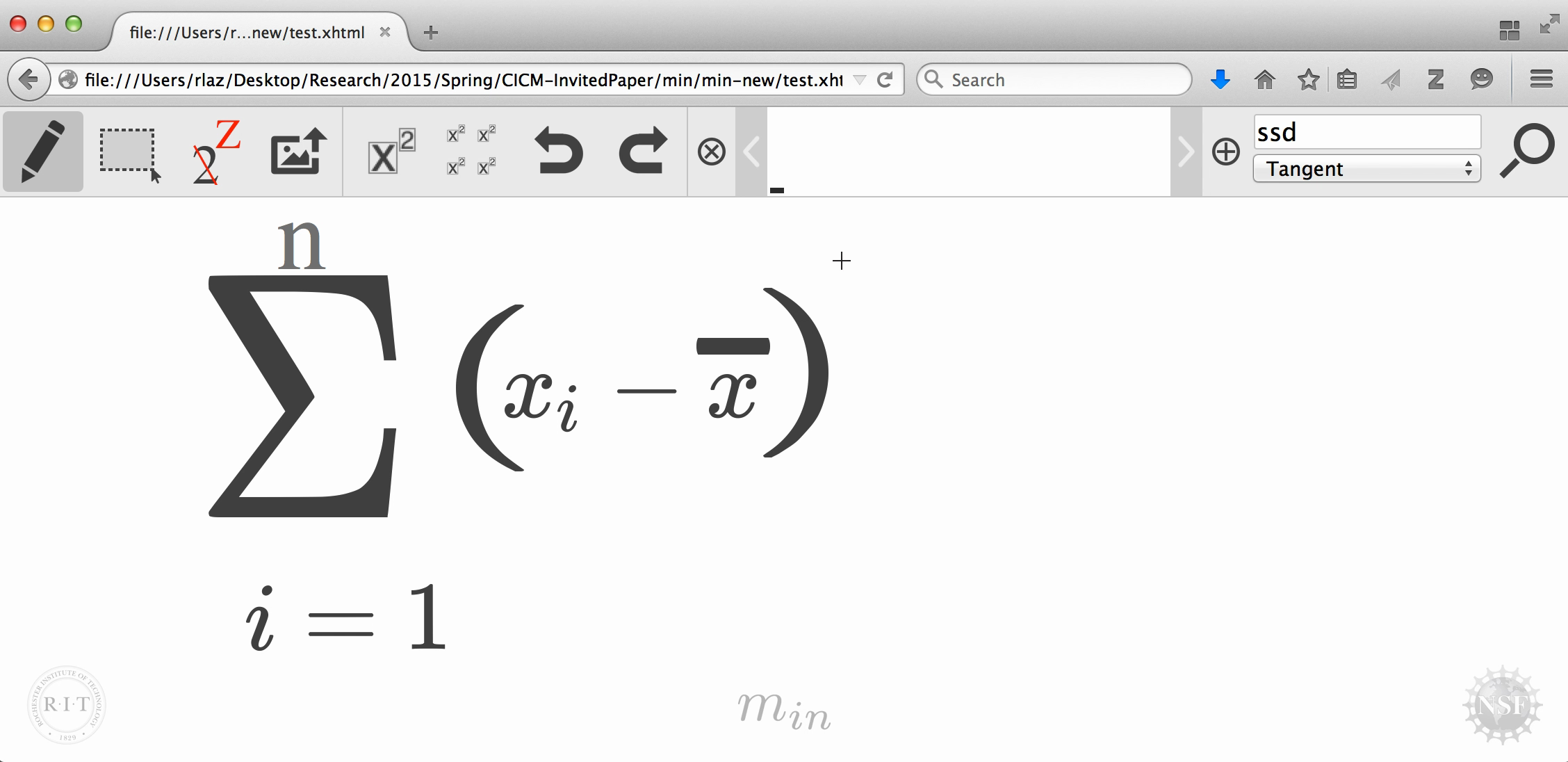}} &

\scalebox{0.04}{\includegraphics{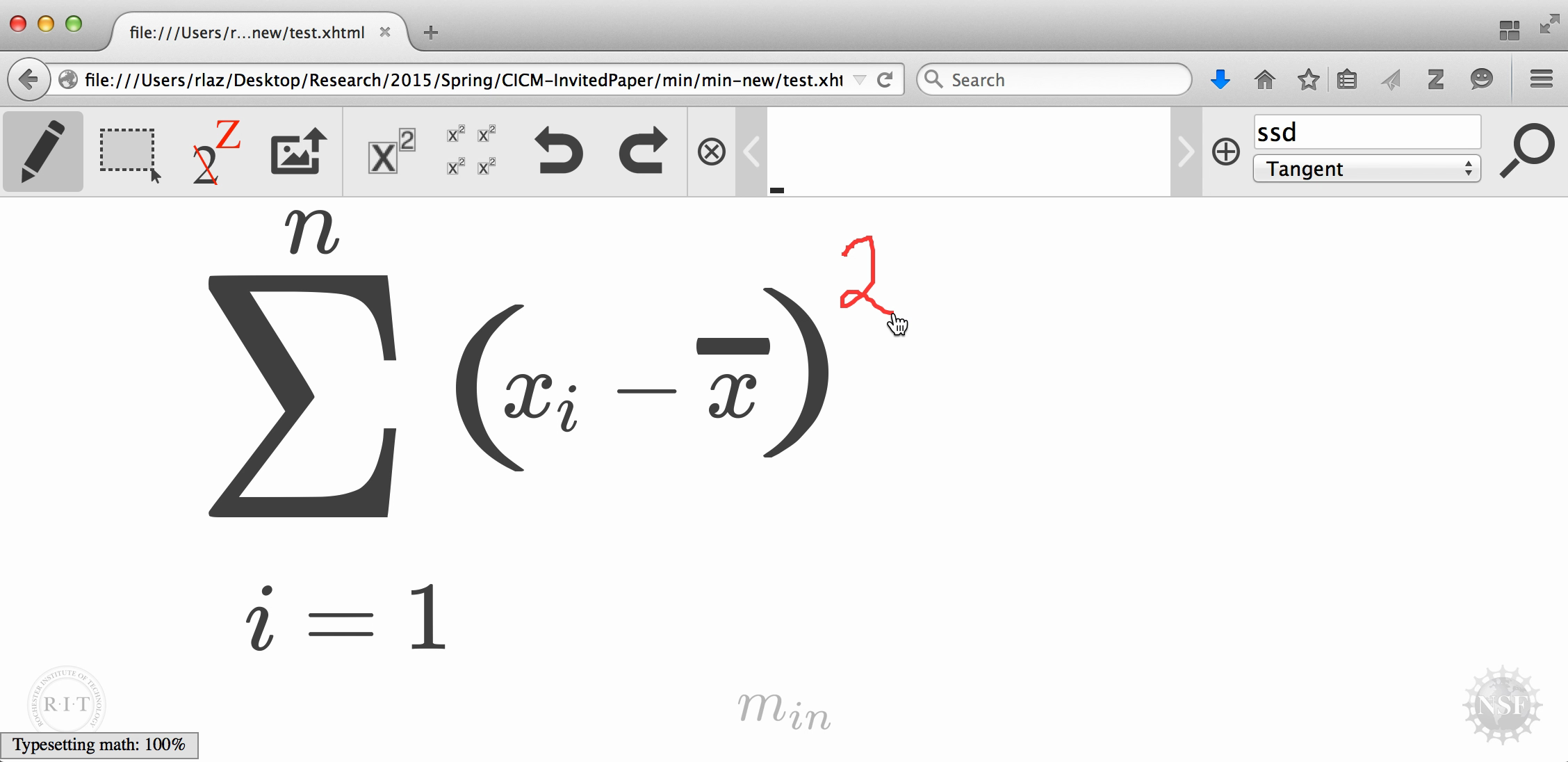}} &
\scalebox{0.04}{\includegraphics{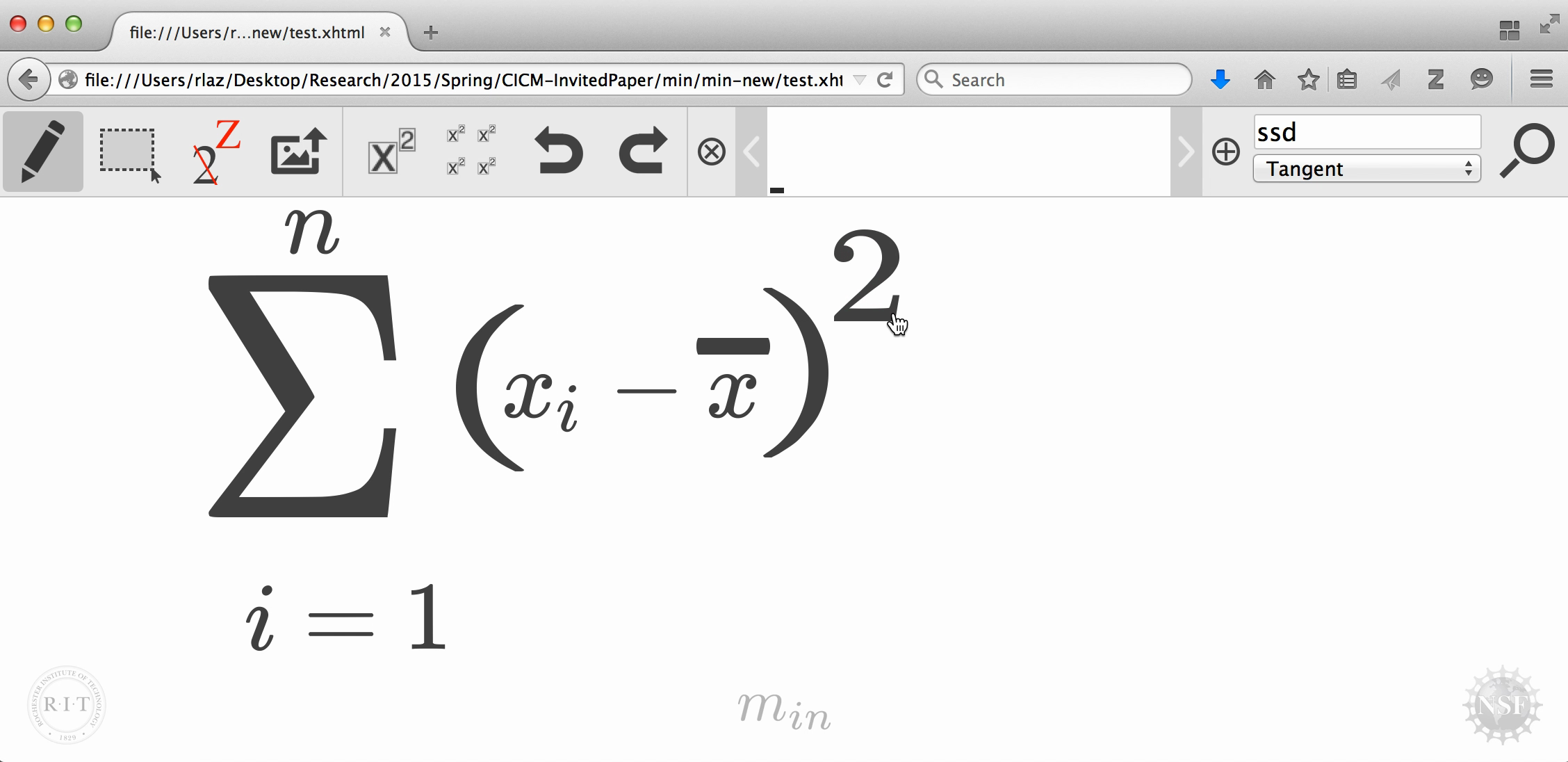}} &
\scalebox{0.04}{\includegraphics{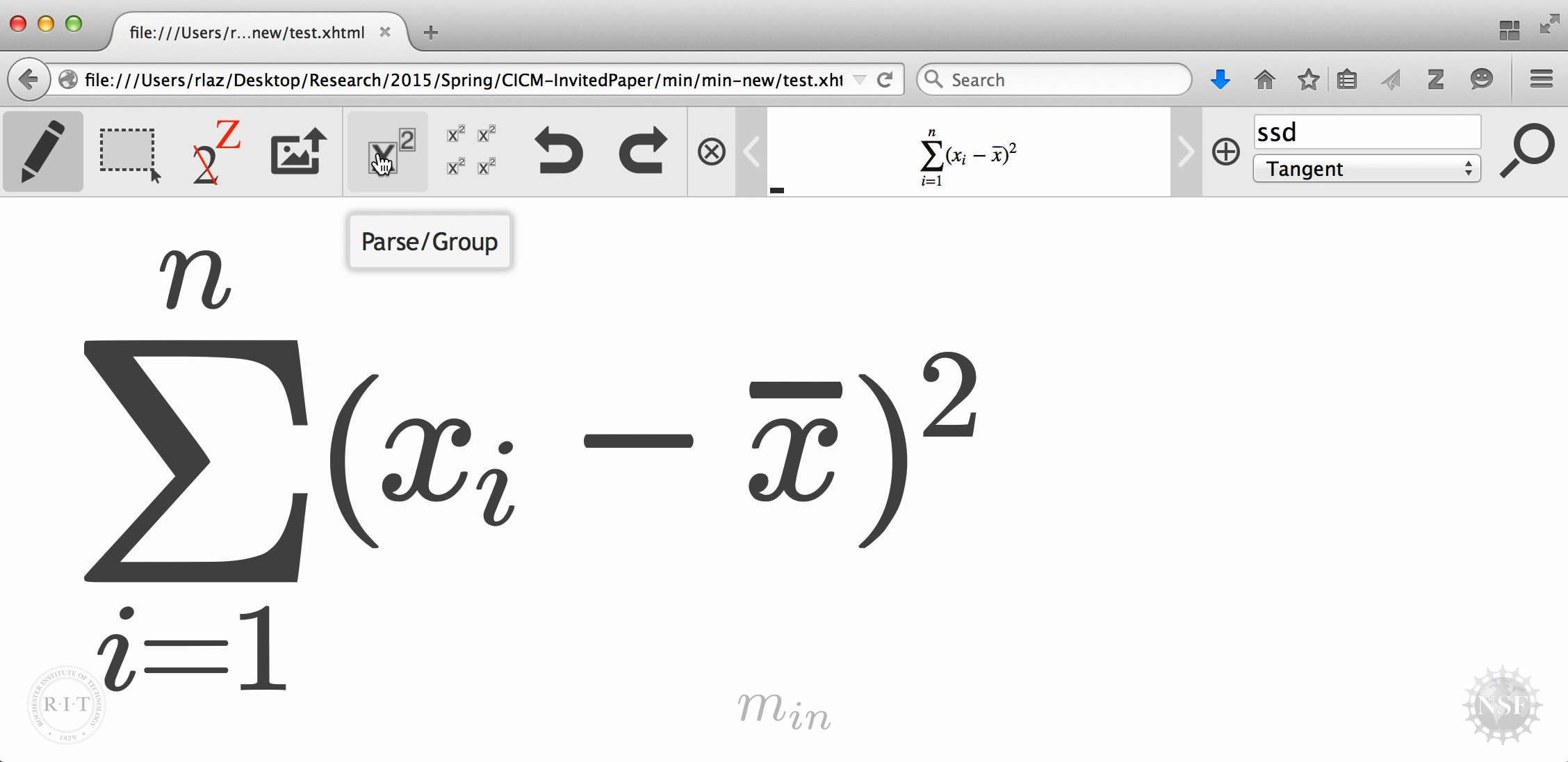}}\\
\end{tabular}\\

\end{center}
\vspace{-0.15in}
\caption{Entering Formula in Figure \ref{fig:min} (from top-left, left-to-right). Symbols are entered through typing and drawing. At bottom-right the `Parse/Group' button is pressed, symbol layout is recognized and symbols are gradually moved (`morphed') to ideal positions}
\end{figure}
\label{fig:entry}
\end{landscape}

Both automatic (solid arrows) and user operations (dotted arrows) are shown in Figure \ref{fig:minarch}.   
Users manipulate symbols through entry, deletion, moving, resizing and merging (e.g. to combine two dashes into an `=' sign). Users also invoke the parser to update the symbol layout tree and produce \LaTeX. When the `Parse/Group' button is pushed by the user, formula structure is visualized two ways, by moving symbols on the canvas to ideal positions, and also by showing the rendered  \LaTeX~in the formula `deck.'

 Currently, operations for entering symbols and recognizing symbol layout are independent of one another. In particular, layout analysis is not performed until explicitly requested by the user. While it is beneficial to integrate classification, segmentation and parsing for automatic recognition \cite{zanibbi2012recognition}, this type of integration may be unhelpful in an interactive system, such as when decisions previously accepted by the user are modified (e.g. if handwritten symbols are re-segmented and re-classified).
Our parser only revises symbols when a compound token is detected, such as replacing a horizontal line with `+' above by `$\pm$.' Our hope is that this behavior is both convenient and predictable.

In the remainder of this section we discuss the recognition modules used in $m_{in}$. The current recognition modules were designed with accuracy, speed, and simplicity in mind. The symbol recognition and parsing modules run externally to $m_{in}$ on servers, with requests made and results transmitted back using simple XML encodings. This allows recognition modules to be easily replaced by services provided on other servers that accept and produce the same encodings. Despite this network overhead, recognition is fast and most first-time users are unaware that recognition is performed remotely.

\subsection{Symbol Entry and Correction}

For formula entry, the grouping (segmentation) of handwritten strokes and symbols in images is performed using simple methods. It is assumed that handwritten symbols are entered one-at-a-time, and recognition is invoked after a short delay (e.g. 1-2 seconds), or when a drawn stroke intersects other strokes (in which case the strokes are merged into a single symbol).  For typeset symbols in images, each separate region of connected black pixels (\emph{connected component}) is treated as a separate symbol. This results in many symbols being over-segmented initially (e.g. `='), but in many cases the DRACULAE parser can locate and correct this by matching rewriting local structures in the symbol layout tree \cite{zanibbi2002recognizing}. A pair of parameters are used to control the location of the centroid used to represent symbol locations, and thresholds to define vertical spatial regions around symbols (above, below, superscipt, subscript).

Handwritten symbol classification is performed  by a Support Vector Machine with a Gaussian kernel applied to modified off-line (i.e. image-based) features \cite{Davila2014fu}. Previously we used Hidden Markov Models \cite{hu2011hmm}. These worked well, but were sensitive to the writing order of strokes. Our new features are insensitive to stroke order and are more accurate as a result. Our classifier is trained using data from the CROHME handwritten math recognition competitions.\footnote{\url{http://www.isical.ac.in/~crohme/}} In the most recent CROHME competition  \cite{Mouchere2014fp} our SVM classifier obtained a test accuracy of 88.7\% for 101 symbol classes, and 83.6\% when invalid symbols are included (102 classes). These rates are within 2-3\% of those obtained by the winning system from MyScript Corporation.\footnote{\url{http://www.myscript.com/}}

Typeset symbols in images (especially digitally-born) tend to be clean and regular, and so we use a simple nearest neighbor classifier. Connected components are assigned to classes using a 10 x 10 histogram of  pixel counts.  We currently use a kd-tree implementation from the Python-based scikit-learn library\footnote{\url{http://scikit-learn.org/}}  for fast approximate nearest-neighbor classification. The classifier is trained using the Infty data sets \cite{uchida2005}. An earlier version obtained recognition rates over 97\% for 190 classes on 70,637 test samples in the \emph{Infty} data set
using pixel histograms \cite{zhu2013rotation}. We do not yet support .pdf input \cite{baker2009,Lin:2014:MRS:2600428.2609611}, but hope to in the future.

In the current version of $m_{in}$, mis-segmented symbols from handwriting or images are deleted and re-entered by users, for example using handwriting or typing.  Both of the handwritten and image-based symbol recognizers return a ranked list of classes that can be selected from the symbol correction menu (see Figure \ref{fig:matrix1}b). This menu also includes a list of symbols organized by type (e.g. digits, latin letters, greek letters, operators, etc.).

\subsection{Parsing Symbol Layout and Generating \LaTeX}

Symbol layout in a formula written on the $m_{in}$ canvas is parsed using DRACULAE \cite{zanibbi2002recognizing},  implemented in the TXL tree rewriting language \cite{Cordy:2006:TST:1149670.1149672}. DRACULAE employs a compiler design, performing a series of tree rewriting passes that: 1) produce an initial symbol layout tree, 2)  replace compound tokens (e.g. replacing two vertically adjacent dashes by `='), 3) rewrite structures such as fractions, and 4) translate the resulting tree to \LaTeX. DRACULAE also produces operator trees where possible (i.e. a `semantic' encoding), but this is unused in $m_{in}$.  In the initial layout analysis step, DRACULAE uses a fast greedy algorithm to recursively locate symbols on the main baseline, and then assigns remaining symbols to regions around baseline symbols (e.g. above, below, superscript, subscripts, within for roots, etc.).

As shown in Figure \ref{fig:matrix1}, users can invoke DRACULAE to parse a subexpression which is then grouped into a unit, `locking' its interpretation \cite{MacLean:2012} and preventing modification by subsequent parses. Symbols and grouped subexpressions may also be combined in a grid, e.g. to enter matrices. This operation uses simple horizontal and vertical bounding box projections to identify gaps for rows and columns - DRACULAE does not recognize matrix structure. Instead, we have DRACULAE treat grouped subexpressions as individual symbols during parsing. Matrix recognition remains a difficult open problem \cite{zanibbi2012recognition,Mouchere2014fp}, but if accuracy can be increased, in the future it may be beneficial to recognize grid cells in addition to rows and columns of predefined cells.

As described earlier, MathJax is used to visualize recognized symbols, and define the ideal locations to which symbols on the canvas are repositioned (morphed) after parsing. 

Parsing errors (e.g. detecting an adjacent symbol as subscripted) are corrected by some combination of moving symbols, undoing the previous parse operation (which `morphs' symbols back to their previous positions), and pressing `Parse/Group' again.



\section{Appearance-Based Math Retrieval}

In this section we summarize a number of different search engines and models designed to support math search using formula appearance. In particular, we describe the \emph{Tangent} search engine and its integration with the $m_{in}$ math search interface, along with methods for visual search of document images and videos.

\subsection{Query-by-Expression for Symbolic Encodings (\LaTeX, MathML)}


Approaches to query-by-expression may be categorized as \emph{text-based} or \emph{tree-based}, as determined by the structures used to represent and retrieve expressions. In text-based approaches, math expressions are linearized before indexing and retrieval. These linearizations are normalized to reduce variability in representation. Common normalizations include defining synonyms for symbols (e.g. function names), using canonical orderings for spatial relationships and commutative operators (e.g. to group `a + b' with `b + a'), enumerating variables, and replacing symbols by their mathematical type.

Linearized math expressions are often handled by term frequency-inverse document frequency-based (TF-IDF) techniques from text retrieval \cite{Miller2003, sojka2011,zanibbi2011keyword}. While linearization loses some formula structure information, it allows text and math retrieval to be carried out in a single framework (usually Lucene\footnote{\url{https://lucene.apache.org/}}).
In a different approach, the largest common string subsequence is used to retrieve \LaTeX~strings
\cite{Kumar2012a}.

Tree-based approaches represent layout or operator trees for formulae directly. 
Methods have been developed that compress tree indices by storing identical subtrees 
in expressions uniquely \cite{Kamali:2010:NMR:1871437.1871635}, with exact matching and tree-edit distances used for retrieval \cite{kamali2013structural}. \emph{Substitution trees} designed for unification have been used to create tree-structured indices \cite{kohlhase2006,schellenberg2012layout}. Descendants of an index tree node contain expressions that unify with the parameterized expression stored at the node (e.g. `$f(~\fbox{1}~)$' unifies with `$f(a)$,' with substitution $\fbox{1} \rightarrow a$). A more recent technique adapts TF-IDF retrieval for vectors of subexpressions and `generalized' subexpressions where constants and variables are represented by a single symbol   \cite{Lin:2014:MRS:2600428.2609611}. Subtrees are normalized for commutative operators and operator precedence, converting symbol layout trees to pseudo-operator trees.

An emerging class of `spectral' tree-based approaches use sets of local structural representations rather then complete subtrees for retrieval. One system converts sub-expressions in operator trees to words representing individual arguments and operator-argument pairs \cite{Nguyen20125820}.
A lattice over the sets of generated words are used to define similarity, and a breadth-first search constructs a neighbor graph traversed during retrieval.  Another system employs an inverted index over paths in operator trees from the root to each operator and operand, using exact matching of paths for retrieval \cite{Hiroya2013yq}. 

Over a number of years our group has developed a novel `spectral' retrieval model, and a search engine implementing the model  \cite{schellenberg2012layout,Stalnaker2015,Pattaniyil2014}. We discuss these in the next section. 

\begin{figure}[!tb]
\begin{center}
\fbox{
	\scalebox{0.3}{\includegraphics{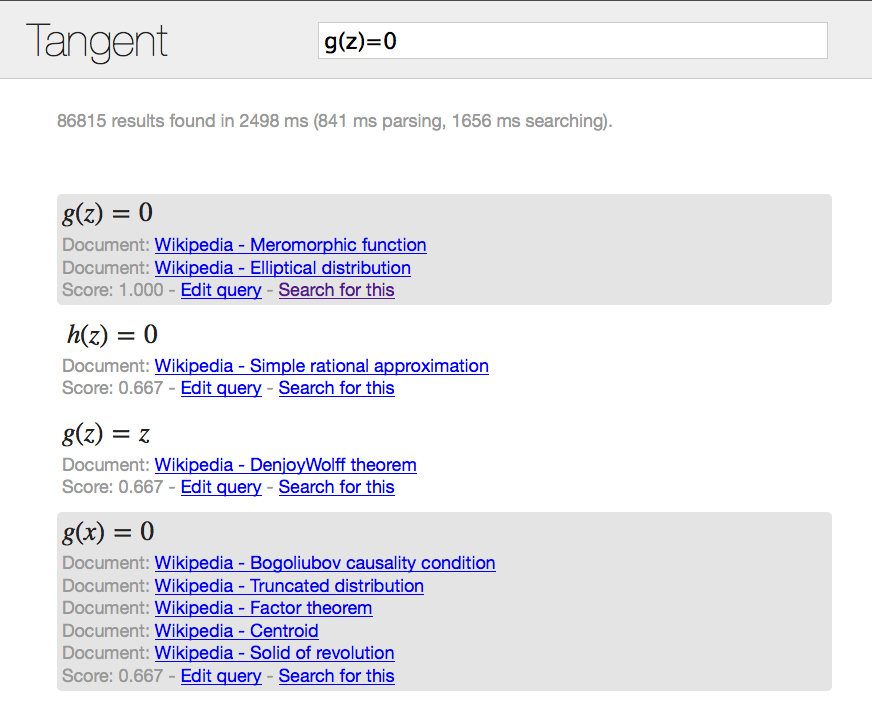}}
}
\end{center}
\caption{Original Tangent Formula Search Engine \cite{Stalnaker2015}. The `Edit query' links send a search hit to the $m_{in}$ search interface for editing and re-submission to Tangent or other search engines. The `Search for this' link supports browsing by allowing hits to be submitted as new queries. Queries maybe typed in \LaTeX~into the text box shown at top, or submitted from $m_{in}$ (see Figure \ref{fig:min}, where Tangent is the selected search engine)}
\label{oldtangent}
\end{figure}

\subsection{The Tangent Math Search Engine}

A screenshot of the \emph{Tangent} search engine\footnote{\url{http://saskatoon.cs.rit.edu/tangent}} is shown in Figure \ref{oldtangent}. The query `$g(z) = 0$' is shown along with the top four matched expressions and their associated Wikipedia articles. The goal with this interface design was to make it easy to use retrieved expressions for editing and search. At the bottom of each hit is a rank score, along with a link to send the hit to $m_{in}$ for editing, and a second link for using a hit to re-query the collection. This integration of $m_{in}$ and Tangent allows for both visual and textual editing of formula queries.

In Figure \ref{oldtangent} we see that Tangent retrieves formulae with structure similar to the query, even when different symbols are used (e.g. `$g$' replaced by `$h$,' `$z$' by `$x$', and `0' by `$z$').  This is interesting, because here only \emph{exact} matching is used for retrieval \cite{Stalnaker2015}. Search results from this first version of Tangent often appear to have performed unification of symbols, but no unification is carried out. This is because the \emph{relative positions of symbols} are used for matching.\footnote{This approach was motivated by a ranking function that used sets of matching symbols and symbol pairs to greatly improve initial retrieval results \cite{schellenberg2012layout}.}

In Figure \ref{oldtangent}, all four hits contain parentheses that are one symbol apart, with an equals sign at right. Matching additional symbol pairs lead to a higher rank. In this example, the first hit is an exact match, with score 1.0 (all symbol pairs are matched), while the remaining three hits have the same rank score, differing by the identity of exactly one symbol relative to the query. This causes relationships with the non-matching symbol to be treated as unmatched. In each case, the five pairs associated with the unmatched symbol are `misses,' out of fifteen total symbol pairs  (for six symbols, ${6 \choose 2} = 15$ symbol pairs, $10/15 = 0.667$).

More concretely, the spectral model used in Tangent represents symbol layout trees by the relative position of each symbol with its descendants in the tree. This is a `bag-of-words' model, with `words' representing relative symbol positions. Note that tuples are not generated for all pairs of symbols when there is branching in the layout tree, unlike the query and hits shown in Figure \ref{oldtangent} which lie on a single baseline. Tuple generation is illustrated in Figure \ref{fig:tangent}a-c. The fraction line has a relationship with every other symbol in the tree, each defined by a pair of integers giving the path length from the line to the symbol in the tree (shown as \emph{Dist.} in Figure \ref{fig:tangent}c, and change in baseline position. The baseline position change is initially 0, increasing by one for each superscript/above relationship and decreasing by one for each subscript/below relationship along the path between two symbols (shown as \emph{Vert.} in Figure \ref{fig:tangent}c \cite{Stalnaker2015}). 

{\bf Tuples in Tangent Version 2 \cite{Pattaniyil2014}.} Later, changes were made to the tuple generation model, adding tuples for symbols in the leaves of layout trees. In Figure \ref{fig:tangent}c, we see that three tuples are defined for the symbols `$2$,' `$y$' and `$z$' at the leaves of the tree shown in Figure \ref{fig:tangent}b. This addition was made to allow single-symbol queries to be represented in the index, particularly to allow matching for matrix subexpressions comprised of a single symbol such as shown in Figure \ref{fig:tangent}d, where three of the matrix entries are a single digit.

A representation for matrices and grid/array structures was also added, such as for expressions shown in Figures \ref{fig:queries}a and \ref{fig:tangent}d. Each grid is represented by a symbol named `matrix' with its dimensions concatenated on the end (e.g. {\bf matrix2x2} in Figure \ref{fig:tangent}f. This `matrix' symbol is then used to represent the entire matrix contents. In Figure \ref{fig:tangent}f the structure of the expression treating the matrix as a unit is contained in rows 6-12 of the table.

Cells/subexpressions in a matrix or grid are represented as independent expressions; in Figure \ref{fig:tangent}f these are the last five rows of the table, representing `$x^2$,' `$0,$' `$0,$' and `$1.$' The subexpression at each matrix location is represented by a tuple giving a row and column location, with the subexpression represented by its \LaTeX~string (as shown in the top five rows of Figure \ref{fig:tangent}f). The idea in this case was to be able to detect when a particular subexpression is present, and also whether the subexpression is located at the correct location in the matrix.

Finally, to support participation in the NTCIR-11 math retrieval tasks \cite{AizKohOun:nmto14,wikipedia}, the Tangent inverted index for tuples was expanded to include entries where one of the two symbols are undefined (e.g. `$x^2$' would be represented concretely, and by `$?^2$' and `$x^?$', where `?' represents a wildcard). Figure \ref{fig:queries}a shows an example of a query containing wildcards. In both tasks, symbols could be replaced by wildcard symbols, which our group interpreted as being any individual symbol.  
Relationships between two wildcard symbols are not indexed, as in some cases will match a vast number of entries in the index (for example, consider `something next to something').

\subsubsection{Retrieval.}

Formula retrieval is performed using an inverted index over symbol pair relationship tuples, mapping tuples to the expressions that contain them. Expressions are represented uniquely, with a separate table recording which documents contain an expression \cite{Pattaniyil2014}.  
 
Queries are first converted to a set of unique tuples with associated counts. Unique tuples are then used to locate matching expressions from the inverted index, and determine the number of instances from the query matched in each retrieved expression. 
Matched expressions are ranked by the harmonic mean of the percentage of pairs matched in the query, and the percentage of pairs matched in the candidate. This may be understood as the f-measure for \emph{recall} of query tuples in the candidate, and  \emph{precision} of tuples in the candidate. This ranking metric prefers larger query set tuple matches, while penalizing unmatched tuples.

In the second version of Tangent, the engine was modified to support both text and multiple formulae in queries. Lucene was used for text retrieval, and formulae were retrieved using Tangent's formula search engine. The formula match score for a document was computed as the sum of the highest formula match scores located for each query expression in a document, each weighted by the relative size of each expression \cite{Pattaniyil2014}. The final rank score for a document was a linear combination of the Lucene-based keyword score and the formula match score. With this, Tangent was now able to handle combined text and formula queries.

\subsubsection{Results.}

A human evaluation compared search results returned by the original Tangent and a Lucene (text-based) retrieval model \cite{Stalnaker2015}. \emph{Precision-at-k} is the percentage of hits in the top-k results deemed `relevant' to a query. In this case, participants were asked to rate hits by their similarity to the query using a 5-point Likert scale from `Very dissimilar' to `Very similar,' with ratings of `Similar' or `Very Similar' treated as relevant. The average precision-at-1 and precision-at-10 values for Tangent were 99\% and 60\%, and 60\% and 39\% for the text-based system. This confirms that using more tree structure produces search results that are perceived as more similar, a result also confirmed in the recent NTCIR math retrieval tasks \cite{AizKohOun:nmto14}.

\begin{figure}[!tb]
\begin{center}
\small
\begin{tabular}{c c}
\hspace{-0.05in}{\scalebox{0.825}{\includegraphics{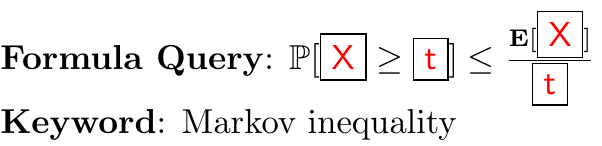}}}
&
\raisebox{1.5em}{

$\mu(A)={\begin{cases}1&{\mbox{ if }}0\in A\\
0&{\mbox{ if }}0\notin A.\end{cases}}$ 
}

\\
a) Math-2 \#39 & b) Wikipedia \#49
\end{tabular}
\caption{Sample Queries from NTCIR-11. Query a) contains four wildcard symbols (shown
in boxes), and two keywords. Queries for the Wikipedia subtask were
single expressions. Query b) has
no wildcards and includes a tabular/matrix layout}
\label{fig:queries}
\end{center}
\end{figure}

\begin{landscape}
\begin{figure}
\centering
   \begin{tabular}{c c c}
  
  {\Huge $ \frac{x^2+y}{\sqrt{z}} $}
   & \raisebox{-5em}{\scalebox{0.4}{\includegraphics{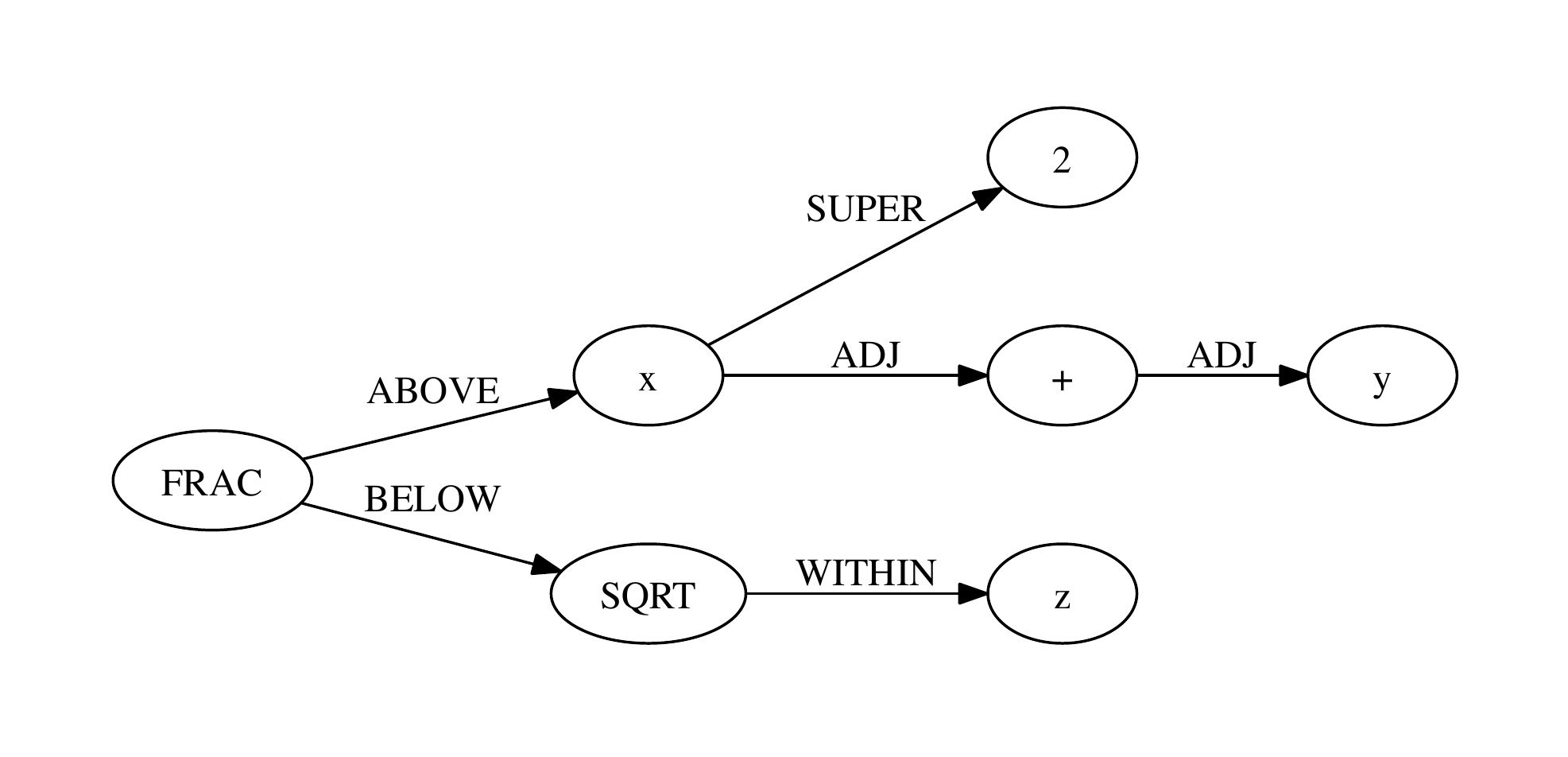}}}   & 
     \raisebox{1em}{ \scriptsize \begin{tabular}{l l r r}
     \hline
        Parent & Child & Dist. & Vert.\\
        \hline
                FRAC & x & 1 & 1       \\
                FRAC & 2 & 2 &2      \\      
                FRAC & + & 3 & 1       \\      
                FRAC & y & 3 & 1 \\
                FRAC & SQRT & 1 & -1  \\
                FRAC & z & 2 & -1 \\
                x & 2 & 1 & 1\\
                    \bf 2 & {\bf None} &\bf  0 & \bf 0 \\

                x & + & 1 & 0\\
                x & y & 2 & 0\\
                + & y & 1 & 0 \\
                            \bf y & {\bf None} & \bf 0 & \bf 0 \\
                SQRT & z & 1 & 0\\
               \bf z & {\bf None} & \bf 0 & \bf 0 \\
                \hline
     \end{tabular}}\\~\\
     a) Expression & b) Symbol Layout Tree & c) Symbol Pair Tuples
%
%
~\\~\\
  
  {\large $\raisebox{0.5em}{A} \begin{bmatrix} x^2 & 0 \\ 0 & 1 \end{bmatrix} \raisebox{0.5em}{+ 1 }$} 
   & \raisebox{-5em}{\scalebox{0.4}{\includegraphics{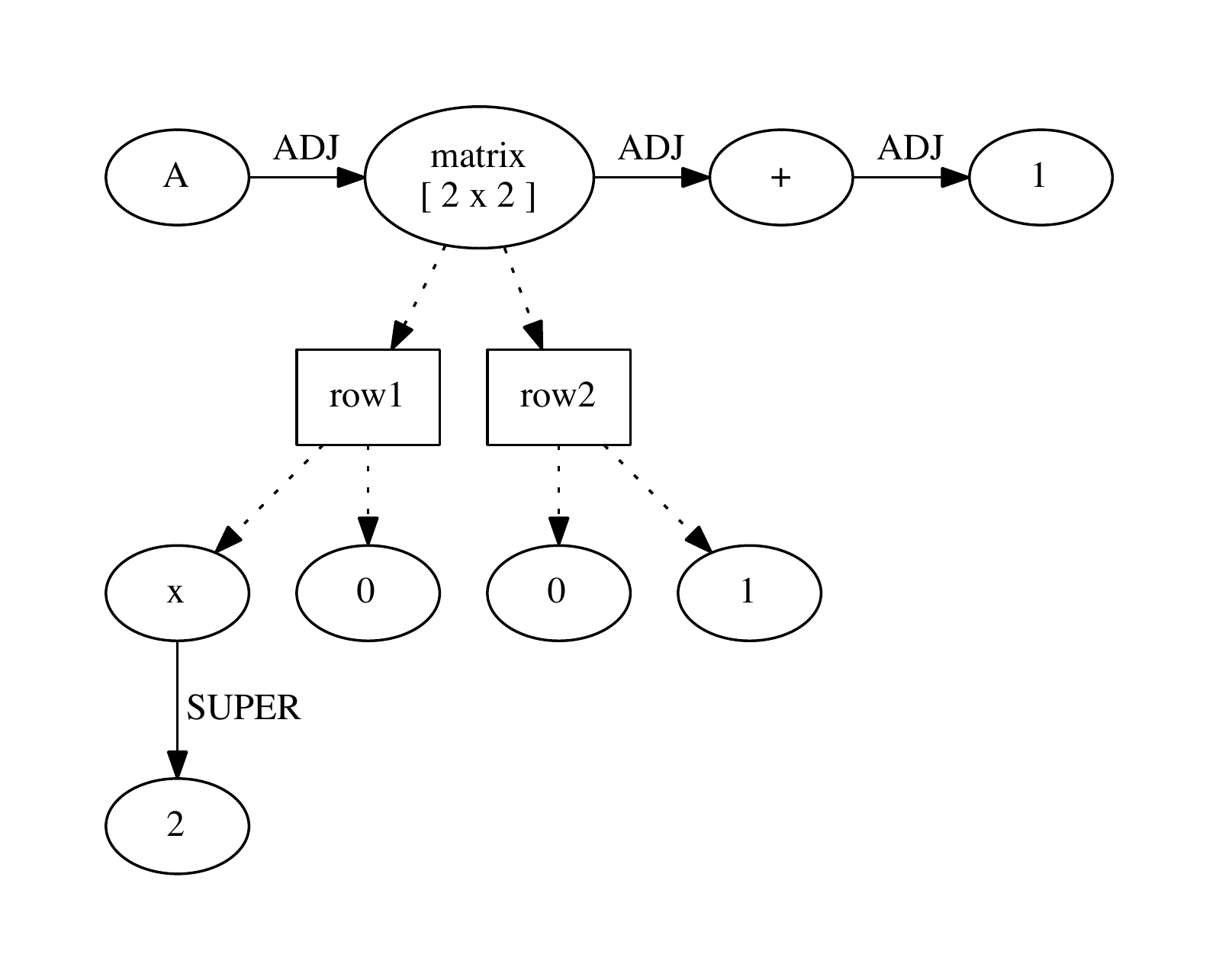}}}   & 
     \raisebox{1em}{ \scriptsize \begin{tabular}{l l r r}
     \hline
    
        \multicolumn{4}{c}{\bf Matrix Structure}\\
            Parent & Child & Row & Column\\
        \hline
        matrix & dimensions & 2 & 2\\
        matrix & `$x^2$' & 1 & 1\\
        matrix & `0' & 1 & 2\\
        matrix & `0' & 2 & 1\\
        matrix & `1' & 2 & 2\\
        \hline
        \multicolumn{4}{c}{\bf Subexpressions}\\
         Parent & Child & Dist. & Vert.\\
         \hline
        A & {\bf matrix2x2} & 1 & 0\\
        A & + & 2 & 0\\
        A & 1 & 3 & 0\\
        {\bf matrix2x2} & + & 1 & 0\\
        {\bf matrix2x2} & 1 & 2 & 0\\
        + & 1 & 1 & 0\\
        1 & None & 0 & 0\\
        \hline
        x & 2 & 1 & 1\\
        2 & None & 0 & 0\\
        0 & None & 0 & 0\\
        0 & None & 0 & 0\\
        1 & None & 0 & 0\\
        \hline
     \end{tabular}}~\\~\\
     d) Expression & e) Symbol Layout Tree & f) Symbol Pair Tuples
    \end{tabular}
   \vspace{0.1in}
   \caption{Tangent: Symbol Pair-Based Layout Representation in for Two Expressions}

    \label{fig:tangent}
\end{figure}
\end{landscape}

The second version Tangent was entered in the NTCIR-11 Math Main Task \cite{AizKohOun:nmto14} and the NTCIR-11 Wikipedia formula retrieval subtask \cite{wikipedia}. Queries from each task are shown in Figure \ref{fig:queries}. The Main task had 50 combined formula and text queries, for a subset of the arXiv containing 100,000 technical papers with substantial mathematics broken up at paragraphs into 8.3 million segments, treated as the documents for the task. Two human evaluators judged hits for the main task to produce precision and related metrics. Tangent produced the highest precision-at-5 measure (92\%), using a 50\%-50\% weighting for combining the text and formula match scores. 

The Wikipedia subtask was a query-by-expression task with 100 queries for 35,000 articles from Wikipedia \cite{wikipedia}. This task used an automated evaluation protocol, ranking system by specific-item retrieval measures (e.g. the rank at which the article from which query expressions were located, and the number of exact matches returned in the top-k hits), without measures for relevance or similarity. For the Wikipedia task, the formula retrieval engine matched matched the highest top-1 score (68\%, obtained by three systems), and overall was amongst the best performing systems in the competition, hampered primarily by queries that contained a large fraction of wildcard symbols (e.g. $\frac{?}{?}$). Considering the manner in which keyword searches are often carried out using a small number of concrete terms, to us it is unclear how frequently queries with a large number of wildcards would be used in a practical setting vs. copying or creating concrete expressions for inclusion in queries. That said, we believe that re-ranking initial results so that variable-variable relationships are not ignored can be used to mitigate this issue.

\begin{figure*}[ptb!]
\centering
\scalebox{0.4}{\includegraphics{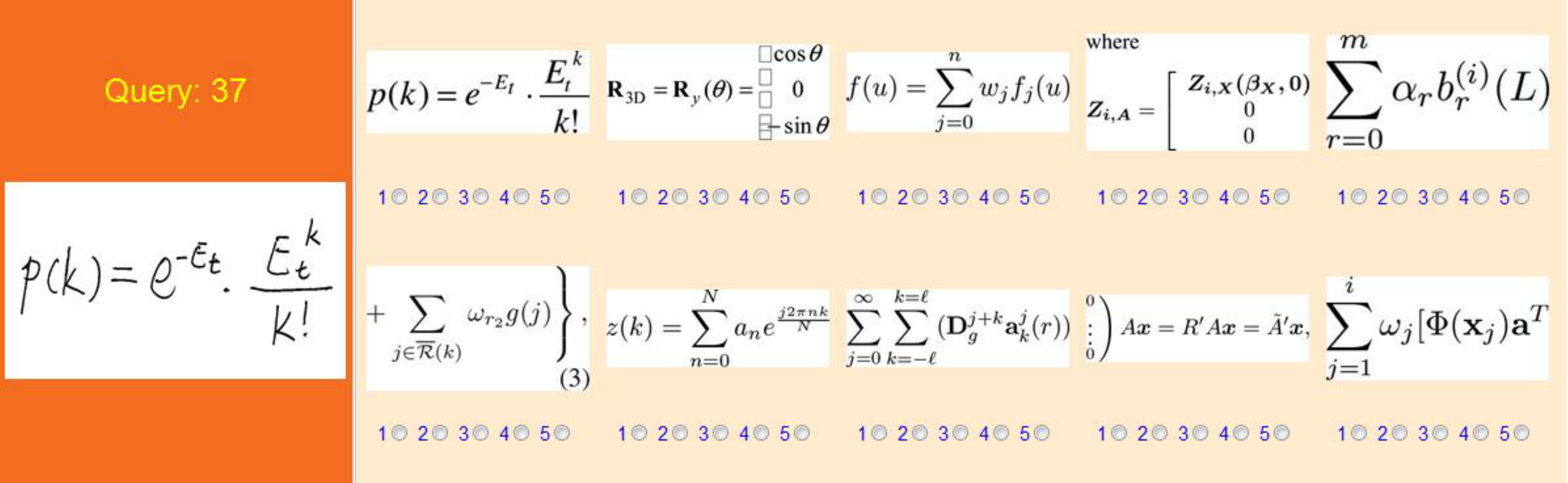}}
\caption{\it User Interface for Evaluating Image-Based Query-by-Expression using Handwritten Queries \cite{zanibbi2011math}}
\label{fig:ui}
\end{figure*}

\subsection{Image-Based Formula Retrieval}

For space we will cover this topic just briefly, but we believe that this is an important future direction for research. Figure \ref{fig:ui} shows an evaluation interface for the first image-based handwritten math retrieval system \cite{zanibbi2011math}. 

In this system layout and contour features measured from an image of a handwritten mathematical expression are used to search document images for similar expressions. Formulae are indexed using X-Y cutting trees \cite{nagy84}, with Dynamic Time Warping of upper and lower image contours used to produce the final ranking (adapting an earlier handwritten word spotting technique \cite{Rath2007}). We were very surprised that our first prototype allowed 10 participants to locate the page from which handwritten queries were taken 63\% of the time in the top 10 on average (20 queries). If the original query images were used, then 90\% of the original queries could be located in the top 10 results.

Related work is currently underway, using image-based retrieval of math in lecture videos using snapshots \cite{Davila2013hq}  and handwritten queries
\cite{DBLP:conf/cec/ChatbriKK14}.

\section{Conclusion: Text + Diagram Search for the Masses}

We have summarized our work on creating interfaces and search engines that support math retrieval using the appearance of mathematical expressions. Our aim in doing this is to help all persons, mathematical non-experts and experts, to retrieve mathematical information naturally using the appearance of expressions, in combination with keywords when appropriate.

A key direction for future research is the creation of intuitive, fast interfaces for diagram copying, editing and inclusion in search queries. $m_{in}$ has made a start in this direction, but much remains to be done. Related to this, we believe that an important future line of research is redefining the conventional text `search box' to include formulae directly. 

Currently, spectral approaches to matching structure in trees appear to be the most promising for appearance-based formula retrieval, such as that used in Tangent. In addition to opportunities defined earlier, identifying ways to reduce index sizes and accelerate retrieval will be important for producing engines that will scale to very large collections, and ideally, internet search engine-scale collections.

In closing, there have been many advances in Mathematical Information Retrieval in recent years, and we believe that progress in searching for diagrammatic notations will dramatically alter the way in which people search for technical information.
It will allow queries to move from ``documents with words similar to \emph{these}"  to also include ``documents with diagrams similar to \emph{these}."

\section*{Acknowledgements}
We thank 
George Nagy, Maria Zemankova, Christian Viard-Gaudin, Harold Mouch\`ere, Frank Tompa and Andrew Kane for helpful discussions. Thanks also to Manish Kanadje for proof-reading a draft of the paper.
This material is based upon work supported by the National Science Foundation (USA) under Grant Numbers IIS-1016815 and HCC-1218801.

%
%

\end{document}